\documentclass{emulateapj}
\usepackage{graphicx}
\usepackage{natbib}

\def\ergss{erg~s$^{-1}$}
\def\LX{$L_X$}
\def\LOIII{$L_{\rm [OIII]}$}

\def\LOIIIcorr{$L_{\rm [OIII],\,corr}$}

\def\fratio{$f_{\rm {H}\alpha}/f_{\rm {H}\beta}$}
\def\lratio{$L_{\rm [OIII]}/L_{X}$}
\def\loglratio{$\log(L_{\rm [OIII]}/L_{X})$}
\def\loglratioc{$\log(L_{\rm [OIII],\,corr}/L_{X})$}

\shorttitle{THE OPTX PROJECT IV}
\shortauthors{Trouille \& Barger}

\begin{document}

\title{THE OPTX PROJECT IV:
How Reliable is [OIII] as a Measure of AGN Activity?
\altaffilmark{1}}

\author{L. Trouille\altaffilmark{2} and A. J. Barger\altaffilmark{2,3,4}}

\altaffiltext{1}{Some of the data presented herein were obtained 
at the W. M. Keck Observatory, which is operated as a scientific 
partnership among the California Institute of Technology, the 
University of California, and the National Aeronautics and Space 
Administration. The observatory was made possible by the generous 
financial support of the W. M. Keck Foundation.}
\altaffiltext{2}{Department of Astronomy, University of Wisconsin-Madison, 
475 N. Charter Street, Madison, WI 53706}
\altaffiltext{3}{Department of Physics and Astronomy, University of 
Hawaii, 2505 Correa Road, Honolulu, HI 96822}
\altaffiltext{4}{Institute for Astronomy, University of Hawaii, 
2680 Woodlawn Drive, Honolulu, HI 96822}

 
\begin{abstract}
We compare optical and hard X-ray identifications of 
active galactic nuclei (AGNs) using a uniformly selected 
(above a flux limit of 
$f_{\rm 2-8~keV}=3.5\times 10^{-15}$~erg~cm$^{-2}$~s$^{-1}$)
and highly optically spectroscopically complete ($>80$\% for
$f_{\rm 2-8~keV}>10^{-14}$~erg~cm$^{-2}$~s$^{-1}$ and
$>60$\% below) $2-8$~keV sample observed in three
{\em Chandra\/} fields (CLANS, CLASXS, and the CDF-N).
We find that empirical emission-line ratio diagnostic 
diagrams misidentify $20-50$\% of the X-ray selected 
AGNs that can be put on these diagrams as star formers, depending on
which division is used. We confirm
that there is a large (2 orders in magnitude) 
dispersion in the log ratio of the [OIII]$\lambda5007$ (hereafter, [OIII])
to hard X-ray
luminosities for the non-broad 
line AGNs, even after applying reddening corrections to the 
[OIII] luminosities. 
We find that the dispersion is similar for the
broad-line AGNs, where there is not expected to be much
X-ray absorption from an obscuring torus around the AGN 
nor much obscuration from the galaxy along the line-of-sight 
if the AGN is aligned with the galaxy.  
We postulate that the X-ray selected AGNs that are misidentified
by the diagnostic diagrams have low [OIII] luminosities due
to the complexity of the structure of the narrow-line region,
which causes many ionizing photons from the AGN not to be
absorbed. This would mean that the [OIII] luminosity can only be used
to predict the X-ray luminosity to within a factor of $\sim 3$ (one
sigma). Despite
selection effects, we show that the shapes and normalizations 
of the [OIII] and transformed hard X-ray luminosity functions 
show reasonable agreement, suggesting that the [OIII] samples 
are not finding substantially more AGNs at low redshifts than 
hard X-ray samples.
\end{abstract}


\section{Introduction}
\label{secintro}

Determining which sources in a large sample of galaxies
host active galactic nuclei (AGNs) at their centers is a 
challenging but essential step for galaxy evolution
studies.  The most common method used optically to separate 
star-forming galaxies from AGNs is a \citet{baldwin81}-type
empirical diagnostic diagram (hereafter, BPT diagram; see 
also \citealt{osterbrock85} and \citealt{veilleux87}) 
of the narrow emission-line ratios [OIII]$\lambda5007$/H$\beta$ 
versus [NII]$\lambda6584$/H$\alpha$.
In this diagram galaxies occupy two well-defined wings:
the left wing consists of star-forming galaxies, while the right 
wing is attributed to galaxies with an active nucleus 
\citep[e.g.,][]{kauffmann03,stasinska06,kewley06}. 
The basic idea underlying these
diagrams is that the emission lines in star-forming galaxies are
powered by massive stars, so there is a well-defined upper limit 
on the intensities of the collisionally excited lines relative
to the recombination lines (such as H$\alpha$ or H$\beta$). In 
contrast, AGNs are powered by a source of far more energetic 
photons, making the collisionally excited lines more
intense relative to the recombination lines.

\citet{kauffmann03} and \citet{heckman04} proposed using the 
luminosity of the [OIII]$\lambda5007$ (hereafter, [OIII]) line 
as a tracer of AGN activity. As they pointed out, the advantages 
of the [OIII] line are that (1) it is typically strong and easy to 
detect; and (2) although it can be excited by both massive stars and 
AGNs, it has been observed to be relatively weak in metal-rich, 
star-forming galaxies.  However, the disadvantage is that one needs 
to assume that [OIII] is an unbiased, orientation-independent 
indicator of the ionizing flux from the AGN.

In the standard `unified' model for AGNs \citep[e.g.,][]{antonucci93}, 
if the observer's view of a galaxy's central supermassive 
black hole and its associated continuum and broad emission-line 
region are unobscured (obscured) by the presence of a dusty torus, 
then it is a type~1 (type~2) AGN.  However, even in the type~2 AGNs 
the covering factor is not one, so radiation is able to escape 
through the opening angle and photoionize the gas located
several hundred parsecs or more from the central engine. 
This gives rise to the narrow emission-line region with
its strong narrow permitted and forbidden emission lines.
Since the narrow emission-line region lies outside of
the dusty torus, the emission lines should not suffer from 
obscuration by that high column density material (though 
they may be affected by dust within the host galaxy).  
Although there is some evidence that the [OIII] luminosity is 
a good measure of AGN activity \citep[e.g.,][]{mulchaey94}, 
and, indeed, many researchers have assumed it to be so in their 
analyses \citep[e.g.,][]{alonso-herrero97,hao05,netzer06,bongiorno10},
recent work has called this conclusion into
question \citep[e.g.,][]{cocchia07,melendez08,diamond-stanic09,lamassa09}.

With the advent of the {\em Chandra\/} and {\em XMM-Newton\/}
X-ray observatories, AGN activity can now also be traced through 
the $2-10$~keV luminosity, although this measurement may suffer from 
some absorption \citep[e.g.,][]{diamond-stanic09,lamassa09}. Also,
X-ray selection misses the most heavily absorbed, Compton-thick sources
with $N_H>10^{24}$~cm$^{-2}$ (see \citealt{comastri04} for a review). 
An interesting question is how well the X-ray 
and optical selections compare.  With our large, uniform, and 
highly optically spectroscopically complete OPTX sample of X-ray 
selected AGNs, we are in a good position to do this comparison.
The OPTX sample consists of one deep pencil-beam survey
(\emph{Chandra} Deep Field-North or CDF-N) and two moderately deep 
wide-field surveys (\emph{Chandra} Large Area Synoptic X-ray Survey or 
CLASXS and \emph{Chandra} Lockman Area North Survey or CLANS). 

In this paper we explore how well the BPT diagram does at 
classifying as AGNs the sources in the OPTX X-ray selected 
AGN sample and how that classification relates to the
very broad observed range in $L_{2-10~{\rm keV}}/$\LOIII\
\citep[e.g.,][]{cocchia07,melendez08,lamassa09}.
Following \citet{heckman05}, \citet{bongiorno10}, and
\citet{georgantopoulos10}, we use our measurements of the mean 
luminosity ratios to transform the fits to the OPTX+{\em SWIFT
BAT\/} hard X-ray luminosity function (LF) from \citet{yencho09} 
and the local {\em RXTE\/} hard X-ray LF from \citet{sazonov04} 
into [OIII] LFs, which 
we then compare with the SDSS [OIII] LFs from \citet{hao05} 
and \citet{reyes08} to investigate the efficiency of optical
versus X-ray selection.

The structure of the paper is as follows.  
In Section~\ref{secxray} 
we briefly describe our $2-8$~keV \emph{Chandra} selected 
OPTX sample.  In Section~\ref{seclines} we determine optical 
narrow emission-line luminosities and investigate the effects 
of line reddening.  In Section~\ref{secbpt} we test how 
many X-ray selected AGNs are misidentified as star formers
by the BPT diagram.
In Section~\ref{seclratio} we plot the ratio of the 
[OIII] luminosities to the $2-8$~keV luminosities versus the
$2-8$~keV luminosities for our non-BLAGNs and 
look for any luminosity dependence, as suggested by
\citet{netzer06}.
In Section~\ref{secdisc} we discuss the probable impact of 
the observed dispersion in this luminosity ratio on the BPT 
classification of X-ray AGNs.  
We then align various hard X-ray and [OIII] LFs for comparison. 
In Section~\ref{secsummary} we summarize our results and 
their implications.

All magnitudes are in the AB magnitude system, and we assume
$\Omega_M=0.3, \Omega_{\Lambda}=0.7$, and H$_0=70$~km~s$^{-1}$
Mpc$^{-1}$.

\section{Sample}
\label{secxray}

In \citet{trouille08} we presented the X-ray catalog for the CLANS
field.  We also presented the optical and infrared photometry and
the spectroscopic and photometric redshifts for the X-ray sources 
in all three of the OPTX fields (CLANS, CLASXS, and CDF-N). 
In \citet{trouille09} we updated the X-ray catalog for
the CLANS field, including 28 sources that had inadvertently
been dropped in the original catalog, and we provided additional 
spectroscopic redshifts for this field. 

In \citet{trouille09} we used a uniform and highly
significant subsample of the OPTX sources to compare the optical
spectral types with the X-ray spectral properties. Here we use the same
selection of sources in the $2-8~\rm keV$ band with significance
greater than $3~\sigma$ and fluxes greater than the flux limits (for a
$S/N=3$) at the pointing center of the $70$~ks CLANS pointings. There
are 746 X-ray sources in this $2-8~\rm keV$ sample (410, 251, and 85
sources from the CLANS, CLASXS, and CDF-N fields, respectively).
Our optical spectroscopic completeness with the DEep Imaging 
Multi-Object Spectrograph (DEIMOS; \citealt{faber03})
on the Keck~II 10~m telescope is very high (see Table~3 of
\citealt{trouille09}), particularly at the bright flux end.
At $f_{2-8~\rm keV}>1.4\times10^{-14}$~erg~cm$^{-2}$~s$^{-1}$ (the break
flux in the $2-8~\rm keV$ number counts; see \citealt{trouille08}) we have
spectroscopic redshifts for 178 of the 217 sources (i.e., we are 82\%
complete to this flux).  

In \citet{trouille09} we followed \citet{szokoly04} and
\citet{barger05} and classified the spectroscopically identified 
sources into four optical spectral types: absorbers (ABS; no strong 
emission lines); (2) star formers (SF; strong Balmer lines and no 
broad or high-ionization lines); (3) high-excitation sources 
(HEX; [NeV], CIV, narrow MgII lines, or strong [OIII]); and 
(4) broad-line AGNs (BLAGNs; optical lines having FWHM line widths 
$>2000$~km~s$^{–1}$). Although the HEX spectral type largely 
overlaps the classical Seyfert~2 spectral type, describing the 
sources as HEX sources helps to avoid confusion with the classical 
definitions.

In \citet{trouille08,trouille09} we used the X-ray fluxes and 
spectroscopic plus photometric redshifts to calculate the 
rest-frame $2-8~\rm keV$ luminosities, $L_X$. At $z < 3$ (which
is all we will be considering in this paper), we 
calculated the luminosities from the observed-frame
$2-8~\rm keV$ fluxes, assuming an intrinsic $\Gamma=1.8$.  That is,
$L_X=f \times 4 \pi d_L^2
\times K-$ correction, where for $z<3$
\begin{equation}
K-\textrm{corr} = (1+z)^{-0.2}~ \textrm{and}~f=f_{2-8~\rm keV} \,.
\end{equation}
Note that using the individual photon indices (rather than the
universal power-law index of $\Gamma=1.8$ adopted here) to calculate
the $K-$ corrections would result in only a small difference in the
rest-frame luminosities \citep{barger02}. We have not corrected the
X-ray luminosities for absorption.  We will examine the impact of this
approach in Section \ref{secdisc}.

Any source more luminous than $L_X=10^{42}$~erg~s$^{-1}$ is very
likely to be an AGN on energetic grounds \citep{zezas98,moran99}. Of
the 746 sources in our $2-8~\rm keV$ sample, 619 have $L_X >
10^{42}$~\ergss. If we adopt the \citet{ranalli03} relation between
star-formation rate (SFR) and X-ray luminosity
\begin{equation}
\label{sfr}
L_{2-10~\rm keV}= \textrm{SFR}/M_{\sun} \times 5 \times 10^{39}~\textrm{erg s}^{-1},
\end{equation}
we 
find that a SFR of at least 200 $M_{\sun}$~yr$^{-1}$ is required
to produce an X-ray luminosity greater than $10^{42}$~\ergss. Given
that the space density of ultraluminous infrared galaxies (ULIRGs, 
which require a SFR $>200~M_{\sun}$) is
approximately $10^{-5}$~Mpc$^{-3}$ at $z\sim 0.9$ \citep{magnelli09},
near the peak in our redshift distribution, and our OPTX
fields cover $\sim1.2$
deg$^{2}$, we expect $\sim 30$ ULIRGs total in our
sample. Because the space density of ULIRGs increases rapidly with
redshift, if instead we restrict our analysis to $z<0.5$ (see
Section~\ref{secbpt}), we expect
fewer than three ULIRGs contaminating our sample. Since the contamination
rate is low ($\sim5$\% of our sample), we assume that the 619 $L_X >
10^{42}$~\ergss\ sources are AGNs (although
see Section~\ref{secbpt} for our classifications of these sources 
as star formation or AGN dominated using emission-line ratio 
diagnostics).  In the remainder of this work, our 
`$2-8~\rm keV$ sample' refers to these 619 sources.

\section{High-Ionization Narrow Emission Lines}
\label{seclines}

\subsection{${\rm [OIII]}\lambda5007$}
\label{oiii}

[OIII] appears in our optical spectra for sources with $z<0.95$. 
Since DEIMOS degrades at the red end, we limit 
our [OIII] analysis to the 157 sources in our $2-8~\rm keV$ 
sample with spectroscopic redshifts less than $0.85$. 
We use the method described in \citet{kakazu07} to determine the
[OIII] flux. We first determine which bandpass covers the redshifted 
emission line. We then integrate the spectrum convolved with that 
bandpass filter response and set the result equal to the broadband 
flux to determine the normalization factor. We
determine the continuum using a sliding 250~\AA\ median. We then multiply the 
integrated continuum-subtracted emission line 
by the normalization factor to obtain the flux in the line. 
This procedure only works for sources with secure continuum
magnitudes where the sky subtraction can be well determined in 
the spectra.  We then convert the emission-line flux into an 
emission-line luminosity using \LOIII~$=f_{\rm [OIII]}\times 
4\pi d_L^2$. 

\begin{figure}[!h]
\epsscale{1.2}
\plotone{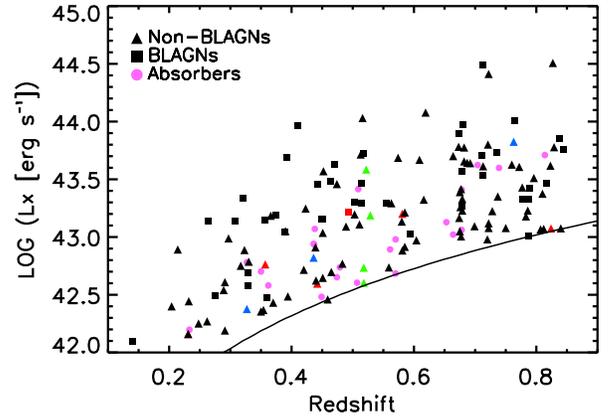} 
\caption{Log$(L_X)$ vs.~redshift for sources in our $2-8~\rm keV$
sample (black---sources with \LOIII~determinations;
magenta---absorbers; red---sources lacking magnitudes in the necessary filter; 
green---sources with sky line interference; blue---sources with 
intervening absorption interference). 
Non-BLAGNs are denoted by triangles, BLAGNs by squares, and
absorbers by circles, as noted in the legend.
The black curve corresponds to the flux limit for our 
$2-8~\rm keV$ sample.
}
\label{LxR}
\end{figure}

In Figure~\ref{LxR} we show rest-frame $2-8~\rm keV$ 
luminosity versus redshift for the sources in our $2-8~\rm keV$ 
sample for which we were able to determine \LOIII~(black symbols;
122 of the 157 sources).  We denote non-BLAGNs by triangles and BLAGNs
by squares. Of the 35 sources that are lacking
\LOIII~determinations, 22 are classified as absorbers
(magenta circles).  By definition, absorbers exhibit little to no
strength in emission lines. However, they do exhibit 
H$\beta$ absorption (and H$\alpha$ as well for those with a low enough
redshift). The presence of these Balmer absorption lines suggests that
they are not BL Lac objects (see \citealt{laurent-muehleisen98} for a
detailed description of BL Lac properties). We note that if these are
in fact low luminosity BL Lac objects, their low non-thermal optical
luminosity could allow for the presence of stellar absorption
lines. We use an equivalent width for the 
[OIII] line that is $1\sigma$ above the noise 
to determine the upper limit on \LOIII~for these sources. 

The remaining 13 sources lack \LOIII~determinations as a result of
observational complications. Six sources (red symbols) are fainter 
than the limiting magnitude in the filter needed to determine 
their \LOIII~values.  Four sources (green symbols) have sky lines that
interfere with the \LOIII~determinations, and three sources (blue symbols) 
have an intervening absorption line located at the position 
of [OIII].  We remove these 13 sources from our subsequent
analysis.

\subsection{Emission-Line Reddening}
\label{secredden}

\begin{figure}[!h]
\epsscale{2.2}
\plottwo{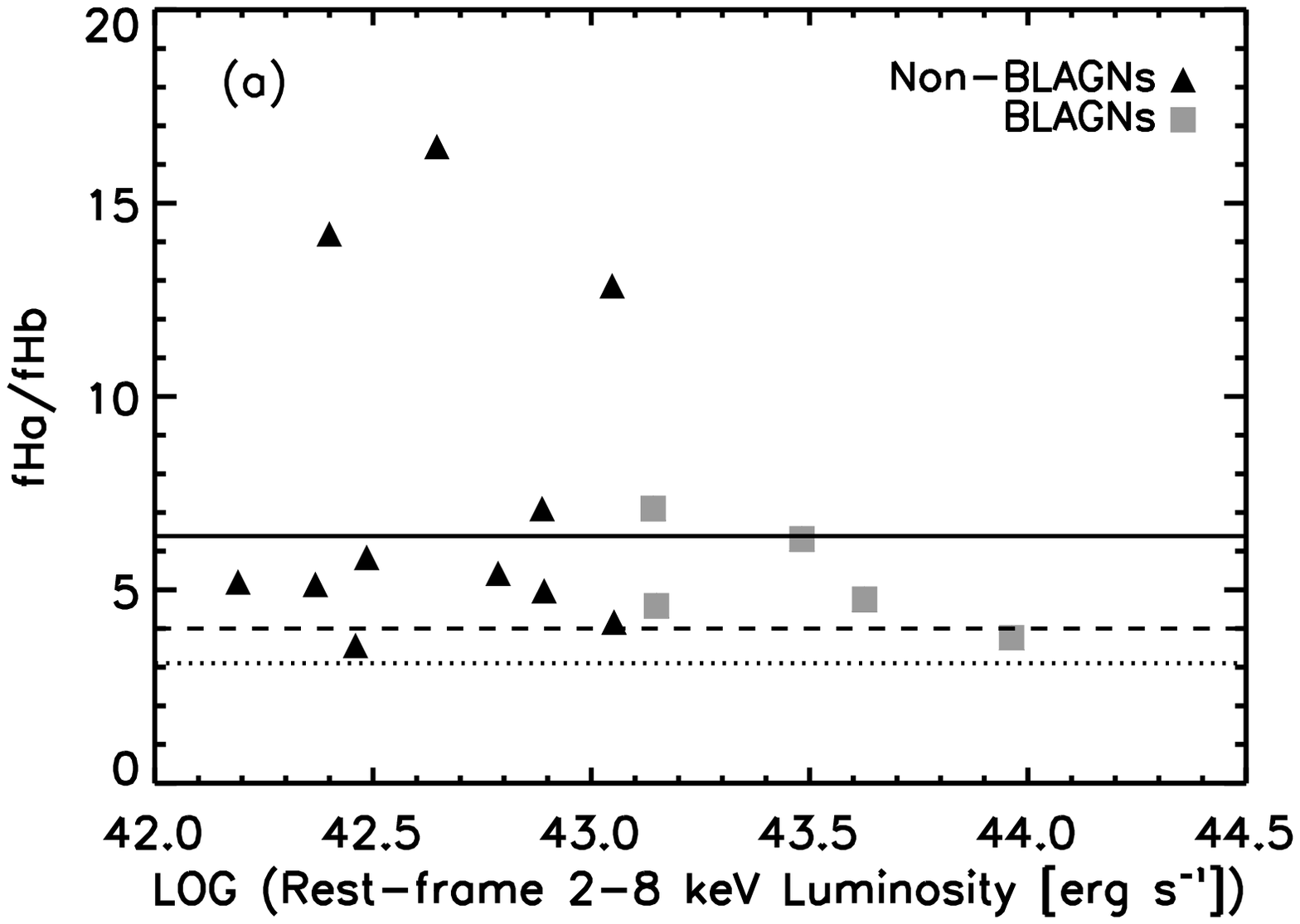}{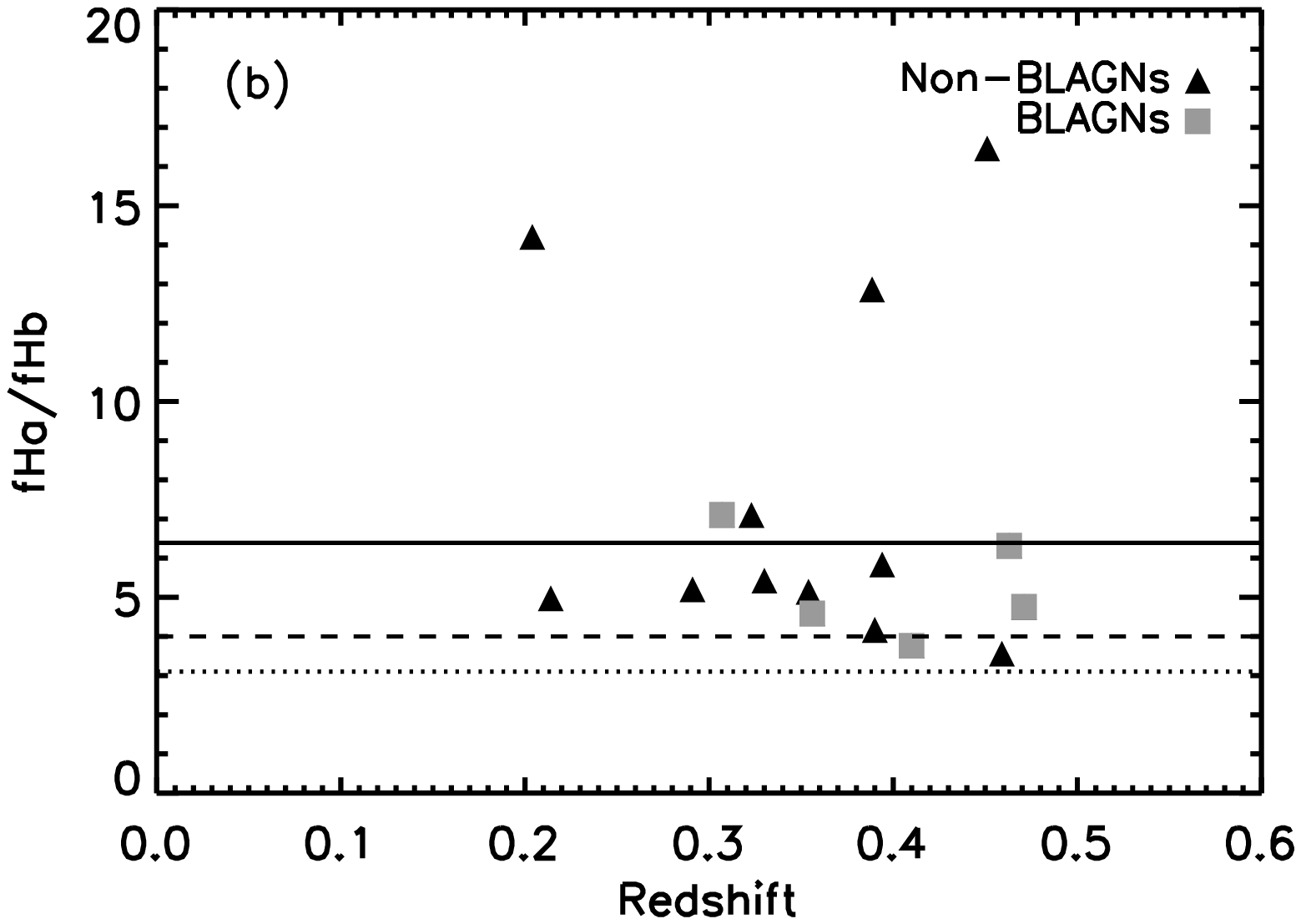}
\caption{(a) Ratio of H$\alpha$ flux to H$\beta$ flux vs.~rest-frame
$2-8~\rm keV$ luminosity for
sources in our $2-8~\rm keV$ sample with reliable $f_{\rm {H}\alpha}$ 
and $f_{\rm {H}\beta}$ measurements (dotted line---the canonical
intrinsic Balmer decrement (\fratio$)_o=3.1$; 
dashed line---the \citet{zakamska03} reddening value for their 
composite SDSS type~2 AGN spectrum, \fratio\ $=4.0$; 
solid line---the \citet{bassani99} median reddening value
for their sample of type~2 AGNs, \fratio\ $=6.4$. 
(b) Same as (a) but versus redshift.}
\label{ratio_z}
\end{figure}

The average reddening towards the narrow-line region in AGNs is still
a matter of debate. Studies using \emph{HST} spectroscopy combined 
with the \emph{IUE}-based analysis of \citet{ferland86} 
suggest that the narrow emission-line reddening for non-BLAGNs
corresponds to $E(B-V) = 0.2-0.4$~mag \citep[e.g.,][]{ferruit99}.
\citet{zakamska03} used the Balmer decrement, \fratio, measured
from a composite spectrum of their SDSS non-BLAGNs as a proxy for the 
volume averaged narrow emission-line reddening and found a similar
result of $E(B-V) = 0.27$~mag (though see \citealt{netzer06} for a 
discussion of possible selection effects). 
\citet{dahari88} and \citet{bassani99} found
significantly larger median reddening in their non-BLAGNs of 
$E(B-V)=0.48$~mag and $E(B-V)=0.65$~mag, respectively. The
\citet{dahari88} sample represented the largest available uniform 
database of optical spectra at that time, including most of the known 
Markarian Seyferts and a majority of the classical Seyferts listed in
\citet{weedman77,weedman78}. The median reddening quoted above is
for the 20 Seyfert~1s in their sample with $z<0.03$ and axial ratios 
$b/a>0.5$. The \citet{bassani99} sample was composed of the 57 Seyfert~2s 
with reliable hard X-ray spectra in the literature at that time. 
They obtained the Balmer decrements from a variety of sources. 
The majority of their sources have $z<0.06$, with a few at $z<0.2$.  

In Figure~\ref{ratio_z} we show \fratio\ versus (a) rest-frame $2-8~\rm keV$ 
luminosity and (b) redshift for the 16 sources in our $2-8~\rm keV$
sample for which 
we were able to measure both the H$\alpha$ and H$\beta$ fluxes. For
reference, there are 45 non-BLAGNs and 19 BLAGNs with $0 < z \le 0.5$ 
(the redshift at which H$\alpha$ 
leaves our spectral window) in our $2-8~\rm keV$ sample. The
separation between BLAGNs (gray squares) and non-BLAGNs 
(black triangles) with luminosity is a result of BLAGNs dominating 
the number densities at higher X-ray luminosities
\citep{steffen03,ueda03,barger05,lafranca05,simpson05,
akylas06,beckmann06,sazonov07,hasinger08,silverman08,winter09,yencho09}.
This result had previously been noted in the radio by 
\citet{lawrence82}.

To determine the H$\alpha$ and H$\beta$ fluxes for the non-BLAGNs we use
the same procedure as described in Section~\ref{oiii} for determining the
[OIII] emission-line fluxes. However, to properly account for stellar
Balmer absorption, we first fit a stellar population model to the
continuum using the Tremonti et al.~special-purpose template fitting
code (private communication), described in detail in
\citet{tremonti04}. In brief, we use a grid of stellar template models
covering a range of age and metallicity 
generated by the \citet{bruzual03} population synthesis code. For each
galaxy we transform the templates to the appropriate redshift and
velocity dispersion to match the data. We then subtract the best-fit
model from the continuum. We only do the subtraction in our
determination of the H$\beta$ emission-line flux, 
because the H$\alpha$
line is typically stronger with a relatively weaker underlying
stellar absorption. We find an average equivalent width of
$\sim1$~\AA\ for the H$\beta$ stellar absorption. This is in
agreement with the fixed offset of $1$~\AA\ used in \citet{cowie08},
which they determined from their averaged absorption-line galaxy
spectrum. 

To determine the H$\alpha$ and H$\beta$ fluxes for the BLAGNs, we fit
both a broad and a narrow component to the lines, subtract the broad
component, and then follow the same procedure as described above. We
are able to determine reliably the H$\alpha$ and
H$\beta$ fluxes for 26\% (22\%) of the BLAGNs (non-BLAGNs) in our
$2-8~\rm keV$ sample with $0 < z \le 0.5$. 

We do not see any statistically significant
difference in the reddening between our BLAGNs and our
non-BLAGNs, nor do we see a dependence on redshift or X-ray
luminosity. We note the small numbers in our
sample and the three non-BLAGNs with significantly higher \fratio\
values. The average \fratio\
for the combined samples is $5.2 \pm 3.1$ 
(corresponding to $E(B-V)= 0.47 \pm 0.4$), which falls
between the \citet[][dashed line]{zakamska03} and 
\citet[][solid line]{bassani99} results. The dotted 
line shows the canonical intrinsic Balmer decrement value,
(\fratio$)_o=3.1$ \citep{ferland83,osterbrock93}. 

Following \citet{bassani99}, we correct our narrow emission 
line fluxes for optical reddening with the equation
\begin{equation}
f_{\rm [OIII],\,corr}=
f_{\rm [OIII]}\times[(f_{\rm {H}\alpha}/f_{\rm {H}\beta})/
(f_{\rm {H}\alpha}/f_{\rm {H}\beta})_o]^{2.94} \,.
\end{equation}
We use our calculated \fratio\ values for the 16 sources in which 
both lines are measured and the average \fratio = 5.2 for the 
remaining sources. This results in an average ratio between our
extinction-corrected and non-extinction-corrected luminosities 
of 4.6.  However, we caution that there remain many uncertainties 
when applying standard extinction corrections to AGN emission-line 
luminosities. For example, \citet{hao05} argue that AGNs 
may have higher intrinsic \fratio\ ratios than normal galaxies
because of higher densities and radiative transfer effects. 
Also, the relative distribution of dust and emission-line gas 
does not necessarily follow the simple model in which dust
forms a homogeneous screen in front of the line-emitting regions.

A Kolmogorov-Smirnov (K-S) test reveals no significant difference
in the distributions of \LOIII\ for our $z \le 0.5$ 
BLAGNs and non-BLAGNs ($p=0.5$). We do not
include the upper limit values for our absorbers in this comparison. 
This is in contrast with \citet{diamond-stanic09}, who
find a statistically significant difference in the
\LOIII\ distributions of their optically-selected Revised
Shapley-Ames sample of Seyfert~1s and Seyfert~2s.


\section{Emission-Line Ratio Diagnostic Diagrams}
\label{secbpt}

BPT emission-line ratio diagnostic diagrams can be used to 
infer whether the gas in a galaxy is primarily being heated by 
star formation or by radiation from 
an accretion disk around a central supermassive black hole. As 
discussed in Section~\ref{secintro}, sources whose signal is 
dominated by accretion lie to the upper-right of those 
dominated by star formation. BPT diagrams are typically 
only applied to narrow-line active galaxies, because in BLAGNs
the narrow lines are overwhelmed by emission from the broad-line 
region. 

\begin{figure}[!h]
\epsscale{1.2}
\plotone{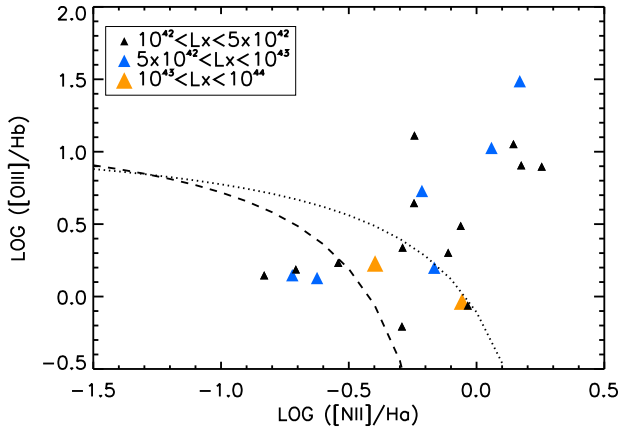}
\caption{(a) A BPT diagram for the non-BLAGNs in our $2-8~\rm keV$
sample with $z \le 0.5$. The symbols grow in size and change in color
with increasing X-ray luminosity, as indicated in the legend. 
The dashed curve denotes the \citet{kauffmann03} division 
between AGNs (upper-right) and star forming galaxies
(lower-left). The dotted curve denotes the \citet{kewley01} division
between AGNs and extreme starbursts.
}
\label{figBPT1}
\end{figure}

In Figure~\ref{figBPT1} we plot $\log($[OIII]/H$\beta)$ versus
$\log($[NII]/H$\alpha)$ for the 21 sources in
our $2-8~\rm keV$ sample for which we were able to
determine the equivalent widths of all four emission lines (note
that all 21 have $L_X>10^{42}$~\ergss). Of the seven $z \le 0.5$
non-BLAGNs with $L_X>10^{43}$~\ergss~and measured [OIII] luminosities, only two
appear in Figure~\ref{figBPT1}. We note that this is a result of
observational complications---sky
lines interfering with our measurement of the H$\beta$, H$\alpha$, or
[NII] lines or because the H$\alpha$/[NII] lines lie within the noise
at the red edge of our spectral window. 

The symbols grow in
size and change in color with increasing X-ray luminosity. 
The dotted curve traces the \citet{kewley01} division between
AGNs and extreme starbursts.  It is a theoretical curve based 
on photoionization models for giant HII regions and a range of 
stellar population synthesis codes.  The dashed curve traces the 
\citet{kauffmann03} division between AGN and star formers, which
is based on an empirical separation of SDSS galaxies.

Less than half ($10/21$) of the X-ray selected sources lie to the upper-right of
the \citet{kewley01} curve and form a sequence similar to that of
the SDSS Seyfert~2s, emerging from the HII region sequence and 
extending to the upper right-hand side of the diagram.
Hereafter, we refer to these sources as `BPT AGN-dominated 
non-BLAGNs'.  We also refer to the 
sources that lie to the lower-left of the \citet{kauffmann03} 
division as `BPT SF-dominated non-BLAGNs' and those
that lie between the two curves as `BPT transition non-BLAGNs'. We
note that of the three BPT SF-dominated non-BLAGNs with $10^{42}< L_X <5\times
10^{42}$~\ergss\ (black triangles), none have
$L_X<2.5\times10^{42}$~\ergss. A SFR of $>500~M_{\sun}$ is required to
produce this level of X-ray luminosity (see Equation \ref{sfr}). Based
on the space density of such strongly starbursting sources,
we do not expect any in our $z<0.5$ $2-8~\rm keV$ sample. 

Since two $L_X>10^{43}$~\ergss~AGN lie in the transition
region and two $5\times10^{42}<L_X<10^{43}$~\ergss~AGNs lie 
below the \citet{kauffmann03} line, it is clear that the 
optical diagnostic diagram is not able to classify all of the 
X-ray selected non-BLAGNs with line ratio measurements as AGNs. Given
that we are working from a 
high X-ray luminosity sample, this seems rather surprising.  
However, since it has been shown by many authors that there 
is considerable spread in \lratio\
\citep[e.g.,][]{alonso-herrero97,bassani99,guainazzi05,heckman05,
ptak06,netzer06,cocchia07,melendez08,diamond-stanic09,lamassa09}, 
it is worth exploring whether this might affect how much overlap 
there is between the two AGN selections.

\begin{figure}[!h]
\epsscale{1.2}
\plotone{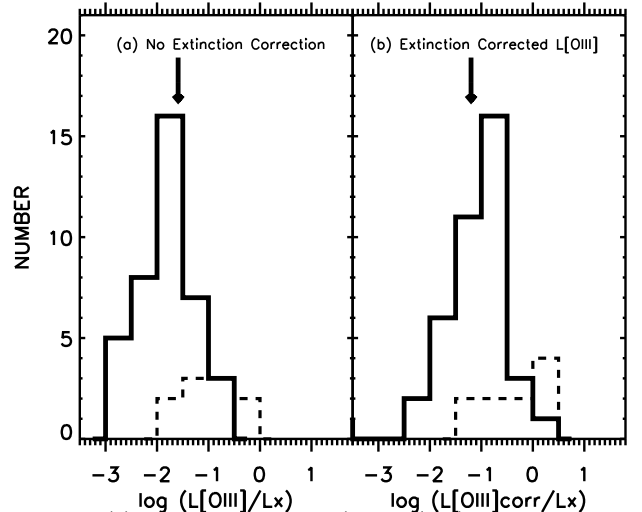}
\caption{(a) Histogram of the \loglratio\ for the sources in our
$2-8~\rm keV$ sample (solid line---$z \le 0.85$
BLAGNs; dashed line---$z \le 0.5$ BPT AGN-dominated non-BLAGNs). 
The arrow marks the mean value of the \citet{heckman05} extinction-uncorrected  
[OIII]-bright optically selected sample of Seyfert~1s. 
(b) Same as (a) but for \loglratioc. The
arrow marks the mean value of the \citet{mulchaey94}
extinction-corrected optically selected
sample of Seyfert~1s.
}
\label{hist}
\end{figure}


\section{The Relation Between \LOIII\ and \LX}
\label{seclratio}

In Figure \ref{hist} we show the distribution of (a) \loglratio\
and (b) \loglratioc\ for the BPT AGN-dominated non-BLAGNs 
(dashed line).  They cover two orders in magnitude, confirming
previous results that showed a very large spread in this ratio.
We also  show the distribution for the $z \le 0.85$ BLAGNs (solid line).  

In Table~\ref{ratio_table} we give the mean and standard
deviation in \loglratio\ and \loglratioc\ for the $z \le 0.5$ BPT 
AGN-dominated non-BLAGNs, for the $z \le 0.5$ and $z \le 0.85$
non-BLAGNs, and for the
$z\le 0.5$ and $z \le 0.85$ BLAGNs. In determining the mean values for
the $z \le 0.5$ and $z \le 0.85$ non-BLAGNs, we do not include the
upper limit values for the absorbers (see Section~\ref{oiii}). 
We find that the mean values for the $z \le 0.85$ BLAGNs agree 
with those from the optically selected \citet{heckman05}
extinction-uncorrected and \citet{mulchaey94} extinction-corrected
Seyfert~1 samples (as indicated by the arrows in Figure \ref{hist}). 

\begin{table}
\caption{$\log(L_{\rm [OIII]}/L_X)$}
\label{ratio_table}
\begin{tabular}{lccc}
\tableline\tableline
   & Mean\tablenotemark{a} & ($\sigma$)\tablenotemark{a} \\
\tableline
BPT AGN-dom non-BLAGNs ($z \le 0.5$) & -1.06 (-0.36)  & 0.6 (0.6) \\
Non-BLAGNs ($z \le 0.5$)\tablenotemark{b} & -1.46 (-0.72) & 0.6 (0.7)\\
Non-BLAGNs ($z \le 0.85$)\tablenotemark{b} & -1.66 (-0.97) & 0.5 (0.6) \\
BLAGNs ($z \le 0.5$) & -1.85 (-1.27)  & 0.5  (0.8) \\
BLAGNs ($z \le 0.85$) & -1.76 (-1.14)  & 0.5  (0.7) \\
\tableline
\tableline
\end{tabular}
\footnotesize
\tablenotetext{a}{The values in parentheses are the \LOIIIcorr~results.}
\tablenotetext{b}{Mean and $\sigma$ do not include absorber upper limit values.}
\end{table}

\begin{figure}[!h]
\epsscale{1.2}
\plotone{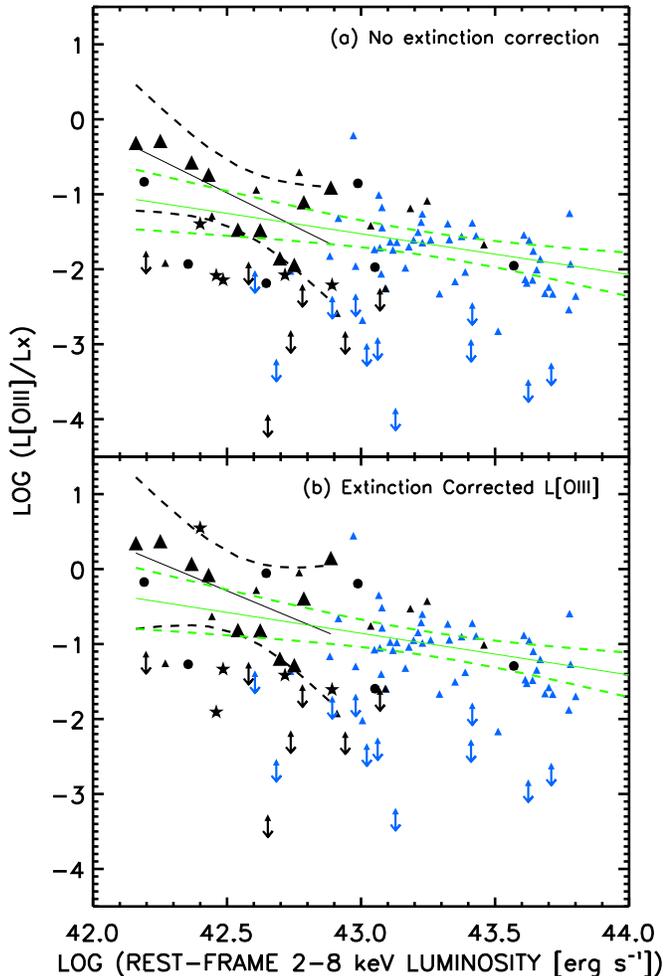}
\caption{(a) \loglratio\ vs.~$\log$~\LX\ for the
$L_X>10^{42}$~erg~s$^{-1}$ non-BLAGNs in
our $2-8~\rm keV$ sample (black small triangles---$z \le 0.5$ non-BLAGNs
lacking BPT line ratio measurements (excluding absorbers); 
black large triangles---$z \le 0.5$ BPT AGN-dominated non-BLAGNs; 
stars---$z \le 0.5$ BPT SF-dominated non-BLAGNs; circles---$z \le
0.5$ BPT transition
non-BLAGNs; blue small triangles---$0.5<z \le 085$ non-BLAGNs). The
upper limits indicate the absorbers in our $2-8~\rm keV$ sample (see
Section~\ref{seclines}). The \LOIII\ values are not extinction-corrected.  The
black (green)
solid line shows the linear best fit to the $z \le 0.5$ BPT
AGN-dominated non-BLAGNs (the $z \le 0.5$ BPT AGN-dominated sources
plus the $0.5 <z
\le 0.85$ non-BLAGNs). The dashed curves show the 95\% confidence 
intervals. We do not include the absorber upper limit values in our
best fit. (b) Same as (a) but for \LOIII\ extinction-corrected values.
\label{ratio_OIII}
}
\end{figure}

\begin{figure}[!h]
\epsscale{1.2}
\plotone{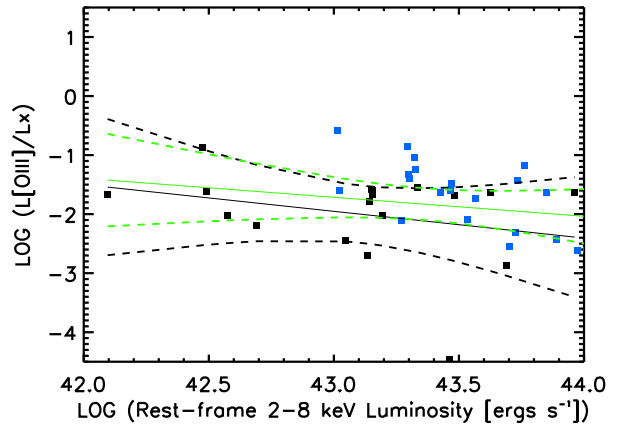}
\caption{\loglratio\ vs.~$\log$~\LX\ for the
$L_X>10^{42}$~erg~s$^{-1}$ BLAGNs with $z \le 0.5$  (black squares)
and $0.5<z \le 0.85$ (blue squares)
in our $2-8~\rm keV$ sample. The \LOIII~values are not
extinction-corrected. The black (green) solid line shows the linear
best fit to the $z \le 0.5$ ($z \le 0.85$) BLAGNs. The dashed curves
show the 95\% confidence intervals. 
\label{ratio_OIII_bl}
}
\end{figure}

In Figure~\ref{ratio_OIII} we plot (a) \loglratio\ versus $\log$~\LX\ 
and (b) \loglratioc\ versus $\log$~\LX\ for the non-BLAGNs 
in our $2-8~\rm keV$ sample.  The black small triangles denote
$z \le 0.5$ non-BLAGNs lacking BPT line ratio measurements (excluding
absorbers), the black large
triangles denote $z \le 0.5$ BPT AGN-dominated non-BLAGNs, the
stars denote $z \le 0.5$ BPT SF-dominated non-BLAGNs, and the 
blue small triangles denote $0.5 < z \le 0.85$ non-BLAGNs. The upper limits
indicate the absorbers in our $2-8~\rm keV$ sample.  

Using a sample of 52 X-ray selected type~2 AGNs from the 
\emph{Chandra} Deep Field-South and HELLAS2XMM,
\citet{netzer06} found that \loglratio\ decreases with
increasing X-ray luminosity 
[linear best fit: \loglratio$=-(0.42\pm 0.07)~\log L_X+(16.5\pm 2.9)$]. 
However, only eight of their sources have direct [OIII]
measurements.  For the others, they determined [OIII] using
transformations from other available narrow emission lines. 
For example, for 32 of their sources they measured the [OII] 
emission line flux and then converted it to [OIII] using the 
mean [OIII]/[OII] flux ratio from \citet{zakamska03}.
They do not take into account any luminosity dependence 
in this ratio, nor is it known whether the \citet{zakamska03} 
mean value applies to an X-ray selected sample.  

In Figure~\ref{ratio_OIII} the black
(green) solid lines show the best fit to the $z \le 0.5$ BPT
AGN-dominated non-BLAGNs ($z \le 0.5$ BPT AGN-dominated non-BLAGNs
plus the $0.5< z \le 0.85$ non-BLAGNs). The dashed curves provide the 95\%
confidence intervals. We do not include the absorber upper limit
values in our best fit.  In Table~\ref{bestfit} we give these best fit
and confidence values.  

Using the Spearman Rank analysis we find a $1.6~\sigma$ 
significance against a null result for the extinction-corrected $z \le
0.5$ BPT AGN-dominated
non-BLAGNs, which indicates little or no 
luminosity dependence. The Pearson coefficient of $-0.6$ 
indicates a weak anti-correlation (a Pearson coefficient of 
$-1$ indicates an anti-correlation, a coefficient of 
0 indicates a null result, and a coefficient of 1 indicates 
a positive correlation). If we include the $0.5 < z \le 0.85$ non-BLAGNs, then
the significance against a null result goes up to $3.3~\sigma$,
although the Pearson coefficient remains at $-0.5$.  The slope for the
$z\le 0.5$ BPT
AGN-dominated non-BLAGNs plus the $0.5< z \le 0.85$ non-BLAGNs is
consistent within the errors with the \citet{netzer06} results. 

In Figure~\ref{ratio_OIII_bl} we plot \loglratio\ versus $\log$~\LX\ 
for the BLAGNs with $z \le 0.5$  (black squares) and $0.5<z \le 0.85$
(blue squares) in our $2-8~\rm keV$ sample. The black
(green) solid line shows the best fit to the $z \le 0.5$ ($z \le
0.85$) BLAGNs. The dashed curves provide the 95\% confidence intervals.  
In Table~\ref{bestfit} we give these best fit and confidence
values. The slopes are consistent
within the errors with the non-BLAGNs. Using the Spearman Rank
analysis we find a $2.1~\sigma$ significance against a null
result for our extinction-corrected $z<0.85$ BLAGNs and a Pearson
coefficient of -0.3, indicative of little or no luminosity
dependence. 

\begin{table}
\centering
\caption{Best fits to the luminosity dependent ratio of $L_{\rm [OIII]}/L_X$}
\label{bestfit}
\begin{tabular}{lcccc}
\tableline\tableline
\multicolumn{5}{l}{$\log(L_{\rm [OIII]}/L_X) = m~\log(L_X)+b$} \\
   & $m$ & $b$ & S-$\sigma$\tablenotemark{a} & P-coeff\tablenotemark{b}\\
\tableline
\multicolumn{5}{l}{BPT AGN-dom non-BLAGNs ($z \le 0.5$):}\\
\LOIII~ & $-1.77\pm0.6$  & $74.1\pm26.5$ & 2.1 & -0.7 \\
\LOIIIcorr~ & $-1.48\pm0.7$  & $62.7\pm31.9$ & 1.6 & -0.6 \\
\tableline
\multicolumn{5}{l}{BPT AGN-dom plus $0.5<z \le 0.85$ non-BLAGNs \tablenotemark{c}:}\\
\LOIII~ & $-0.54\pm0.1$  & $21.9\pm5.7$ & 3.3 & -0.5 \\
\LOIIIcorr~ & $-0.60\pm0.1$  & $23.0\pm5.9$ & 3.3 & -0.5 \\
\tableline
\multicolumn{5}{l}{BLAGNs ($z \le 0.5$):}\\
\LOIII~ & $-0.45\pm0.4$  & $17.5\pm16.6$ & 0.4 & -0.3 \\
\LOIIIcorr~ & $-0.55\pm0.4$  & $22.4\pm16.7$ & 1.3 & -0.3 \\
\tableline
\multicolumn{5}{l}{BLAGNs ($z \le 0.85$):}\\
\LOIII~ & $-0.32\pm0.2$  & $12.1\pm9.9$ & 1.5 & -0.2 \\
\LOIIIcorr~ & $-0.40\pm0.2$  & $16.1\pm9.8$ & 2.1 & -0.3 \\
\tableline\tableline
\end{tabular}
\tablenotetext{a}{Spearman-Rank analysis.}
\tablenotetext{b}{Pearson coefficient.}
\tablenotetext{c}{Best fit does not include absorber upper limit values.}
\end{table} 

\section{Discussion}
\label{secdisc}

From our sample, which is a high \LX\ 
sample, we saw that there is substantial spread in \lratio\ 
at a given \LX\ for the $z \le 0.5$ BPT 
AGN-dominated non-BLAGNs (see black large triangles in
Figure~\ref{ratio_OIII}), which 
means the sources can scatter all the way from the AGN
portion of the BPT diagram to the star formation portion.
This means that AGNs at the low end of the sample in \lratio\ could be
misidentified as star formers with the BPT diagram. 

\begin{figure}[!h]
\epsscale{1.2}
\plotone{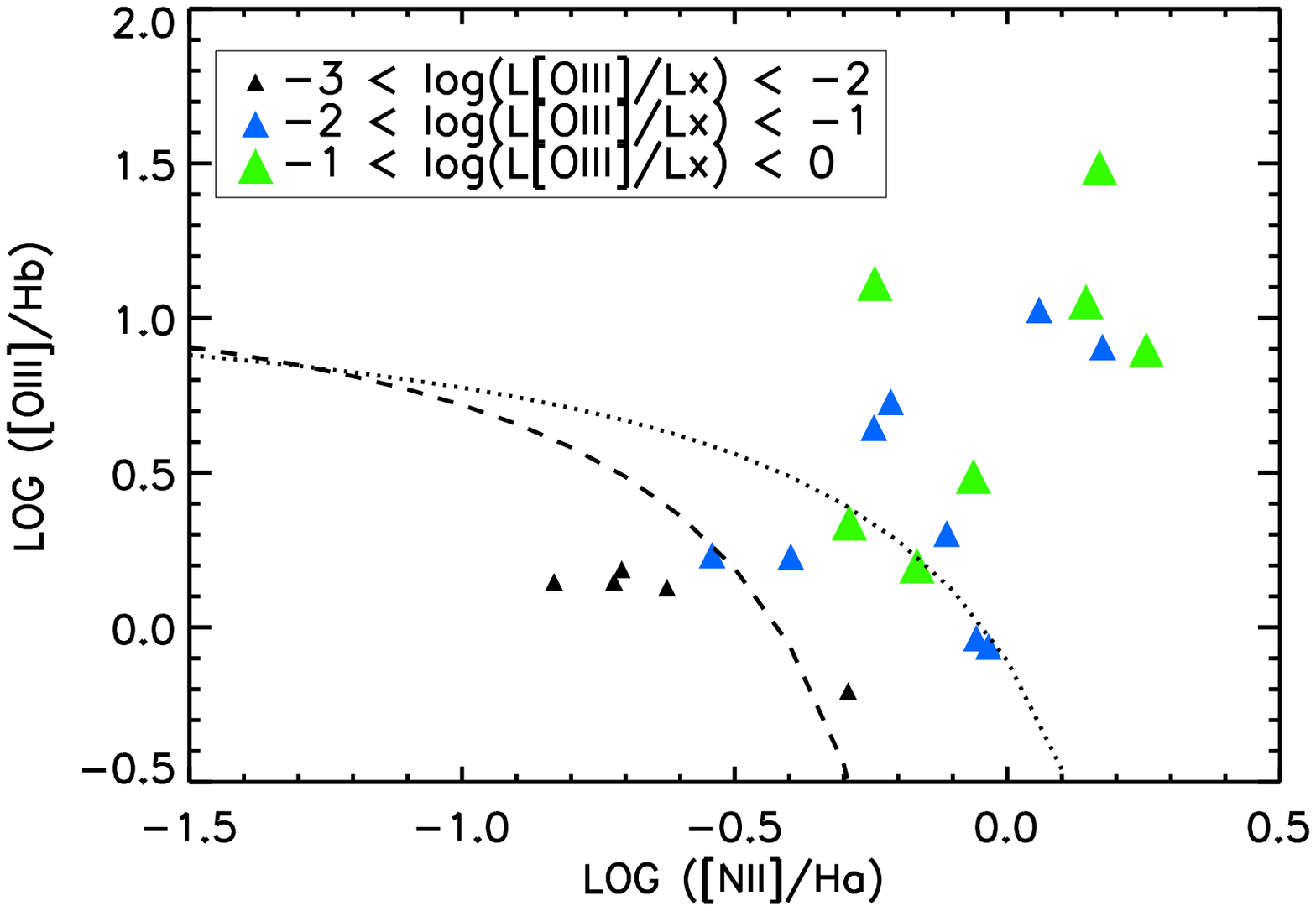}
\caption{A BPT diagram for the non-BLAGNs in our $2-8~\rm keV$
sample with $z \le 0.5$. The symbols grow in size and change in color
with increasing \LOIII/\LX, as indicated in the legend.  
The dashed curve denotes the \citet{kauffmann03} division 
between AGNs (upper-right) and star forming galaxies
(lower-left). The dotted curve denotes the \citet{kewley01} 
division between AGNs and extreme starbursts. 
}
\label{figBPT2}
\end{figure}

In Figure~\ref{figBPT2} we replot the BPT diagram of 
Figure~\ref{figBPT1}, this time using the color and size of
the symbols to indicate an increasing \loglratio\ rather
than an increasing \LX.  It does appear that there is
a trend in \loglratio, with the $z \le 0.5$ BPT SF-dominated 
non-BLAGNs exhibiting the lowest values ($-3<$\loglratio$<-2$). This
is consistent with the idea that the [OIII] emission from some 
AGNs is low.  This dispersion in the amount of [OIII] light 
produced by an AGN also implies that we will
not be able to tell the strength of the AGN relative to
the star formation as we move along the wing, as some
modelers have proposed \citep[e.g.,][]{kewley06}.  Instead
the location along the wing will be a mixture of the AGN 
strength and how much \LOIII\ there is for the AGN strength.

\begin{figure}[!h]
\epsscale{1.2}
\plotone{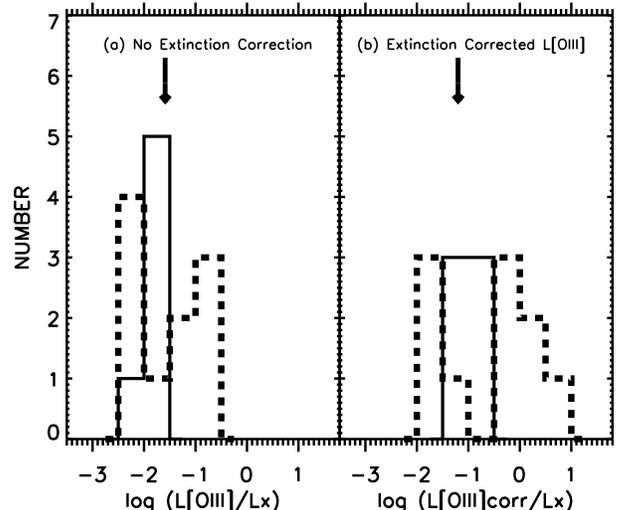}
\caption{(a) Histogram of the \loglratio\ for the $z \le 0.5$ sources in our
$2-8~\rm keV$ sample with reliable H$\alpha$ and H$\beta$ flux
measurements (solid line--- BLAGNs; dashed line--- non-BLAGNs). 
The arrow marks the mean value of the \citet{heckman05}
extinction-uncorrected [OIII]-bright optically selected sample of Seyfert~1s. 
(b) Same as (a) but for \loglratioc. The arrow marks the mean value of
the \citet{mulchaey94} extinction-corrected optically selected
sample of Seyfert~1s.
}
\label{hist2}
\end{figure}

One possible explanation for the low [OIII] emission in
some AGNs is extinction from the host galaxy.
Indeed, in Figure~\ref{ratio_OIII} we see that one of the 
five BPT SF-dominated non-BLAGNs (stars) significantly 
increases its \loglratio\ value as a result of the extinction 
correction. However, the other four retain 
significantly lower \loglratio\ values than the mean for 
the BPT AGN-dominated non-BLAGNs. We note that all five BPT
SF-dominated non-BLAGNs have reliable H$\alpha$ and H$\beta$ 
flux measurements and therefore unique extinction
corrections. The small extinction corrections suggest that host galaxy 
obscuration is not the dominant reason for the low 
[OIII] emission in the majority of our BPT SF-dominated non-BLAGNs.

Figure \ref{hist2} also provides support for the idea that applying
extinction corrections does not help in reducing the scatter in the
ratio between \LOIII~and \LX. In this figure we show the distribution of (a)
\loglratio~and (b) \loglratioc~for the $z \le 0.5$ BLAGNs in our
$2-8~\rm keV$ sample (solid line). We also show the
distribution for our $z \le 0.5$ non-BLAGNs (dashed line). We only include
sources with reliable H$\alpha$ and H$\beta$ flux measurements. While
the numbers are small, we see that correcting for
extinction does not reduce the dispersion. 

It is possible that the extinction corrections determined
from the Balmer line ratios in Section~\ref{secredden}
are not totally correct for the [OIII] emission.
For example, there may be additional contributions to the 
Balmer emission lines from star formation further out in 
the galaxy, so if the dust were not mixed uniformly
throughout, the extinction corrections measured from
the Balmer line ratios might be less than what they should 
be for the narrow-line region.  However, at least
the extinction corrections from the Balmer line ratios
give some impression of how extinction might alter things,
and we see that it does not solve the problem.
Another issue is whether X-ray absorption due to an
obscuring torus around the AGN could be affecting the
luminosity ratio and whether correcting \LX\ for
absorption by the torus would reduce the dispersion. 

\begin{figure}[!h]
\epsscale{1.2}
\plotone{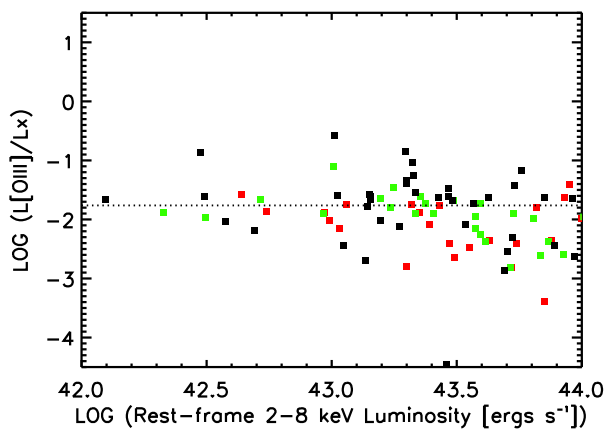}
\caption{\loglratio\ vs.~$\log$~\LX\ for the 
$L_X>10^{42}$~erg~s$^{-1}$ BLAGNs with $z \le 0.85$ in
our $2-8~\rm keV$ sample (black squares). The horizontal dotted line 
denotes the mean $\langle$\loglratio$\rangle=-1.76$ for these sources.
The \LOIII~values are not extinction-corrected. The green squares show
the \citet{heckman05} sample of $z<0.2$ Seyfert~1s from the
\emph{RXTE} slew survey \citep{sazonov04}. The
red squares show the \citet{cocchia07} Seyfert 1 sample from the
HELLAS2XMM survey.
\label{ratio_OIII_bl_comp}
}
\end{figure}

We address these issues in
Figure~\ref{ratio_OIII_bl_comp}, where we show \loglratio\ versus 
$\log$~\LX\ for the BLAGNs in our $2-8$~keV sample
as well as for the BLAGNs in other X-ray selected 
surveys from the literature.  We see that a similar high 
dispersion is also observed for the BLAGNs, even though X-ray 
absorption would be small and extinction from the galaxy should 
not be very severe if the AGN is aligned with the disk of the galaxy.  

We conclude that sometimes [OIII] will diagnose the presence of an AGN
and sometimes it will not, and extinction 
is not the primary factor.  Furthermore, we find that the dispersion
in \lratio\ is larger than the dispersion in $L_{\rm opt}/L_X$, which suggests
that \LOIII\ is particularly problematic. We characterize the optical
continuum luminosity, $L_{\rm opt}$, in our sources by their rest-frame $g$-band
luminosity, $L_g$. We determine $L_g$ by transforming the observed
$g,r,i,z,J,H,K_s$ OPTX photometry \citep{trouille08} into $g$-band
flux at $z=0$ using kcorrect v4\_1\_4 \citep{blanton07}. Since
the contribution from the emission lines in AGNs is so low compared to the
continuum luminosity (see \citealt{heckman04} in which $\langle L_{\rm
  opt}/L_{\rm [OIII]} \rangle \sim 320$ for SDSS type 1 AGNs), $L_g$ provides
a reasonable proxy for
the optical continuum luminosity in these sources. We find a mean
$\log(L_{\rm opt}/L_X)$ for our BLAGNs of 0.94, with a dispersion of 0.36. This
is smaller than the dispersion in \loglratio\ ($\sigma = 0.5$
for our BLAGNs, see Table \ref{ratio_table}). 

Instead we suggest that there is substantial complexity in the structure of the 
narrow-line region, which causes many ionizing photons 
from the AGN not to be absorbed.  This complexity could include
a low covering fraction of material, the configuration of the 
HII regions, the presence of density-bounded HII regions, and 
wind venting.  Hence, \lratio\ would be low, and one
would perceive the source as a star former rather than as an
AGN.

Several groups have compared the hard X-ray and [OIII]
luminosity functions (LFs) for AGNs in the local universe
\citep[e.g.,][]{heckman05,georgantopoulos10} to try to 
determine which selection of AGN is most efficient.  As we 
have shown, the large dispersion in \lratio\ means that
this transformation cannot be done to better than a 
factor of $\sim 3$ (one sigma) for an individual object.  However, even though 
we cannot be confident about the tranformation, it is still 
interesting to explore whether the two LFs have, on average, 
the same shape.

\begin{figure*}[!h]
\epsscale{0.6}
\plotone{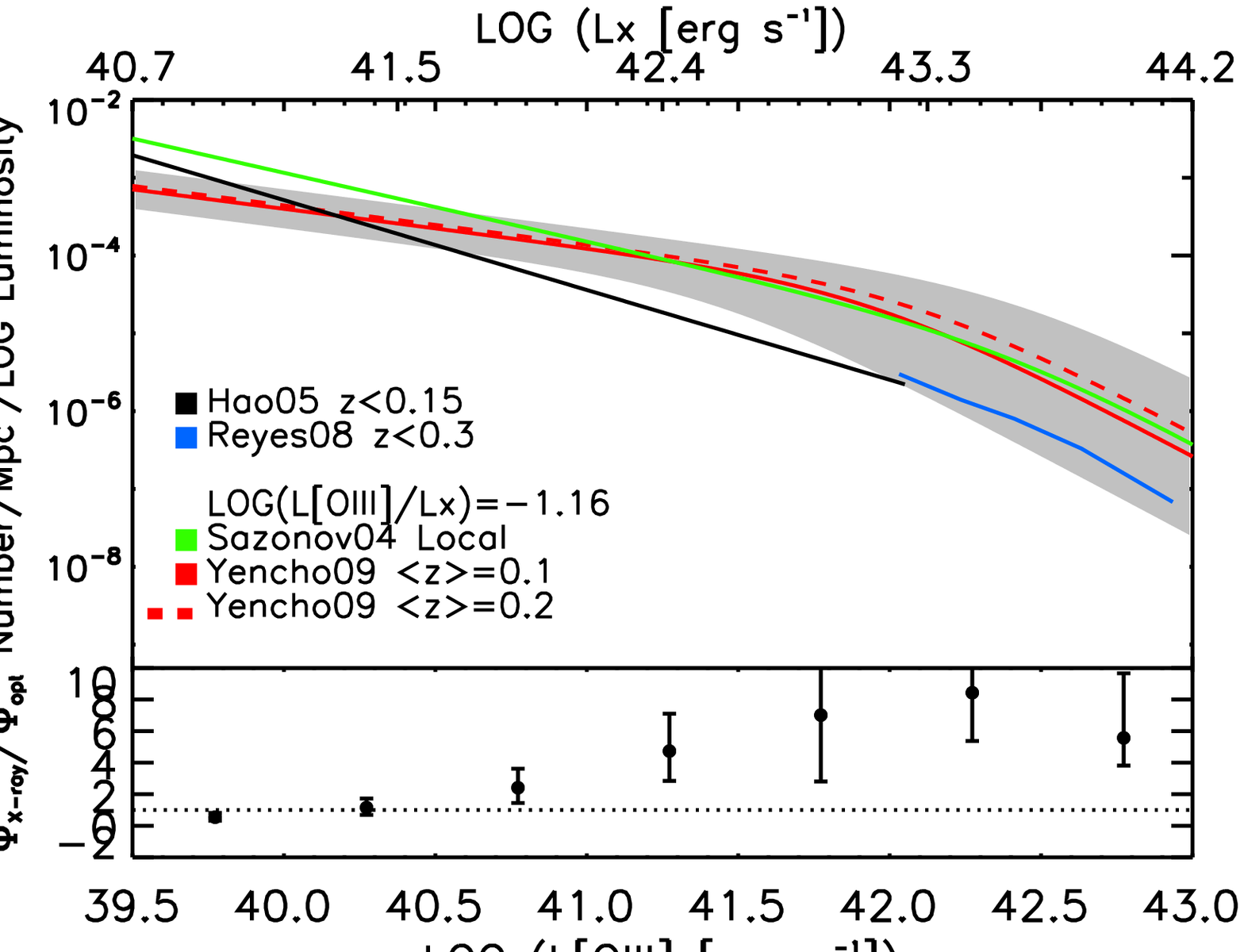} 
\vskip 0.1cm
\plotone{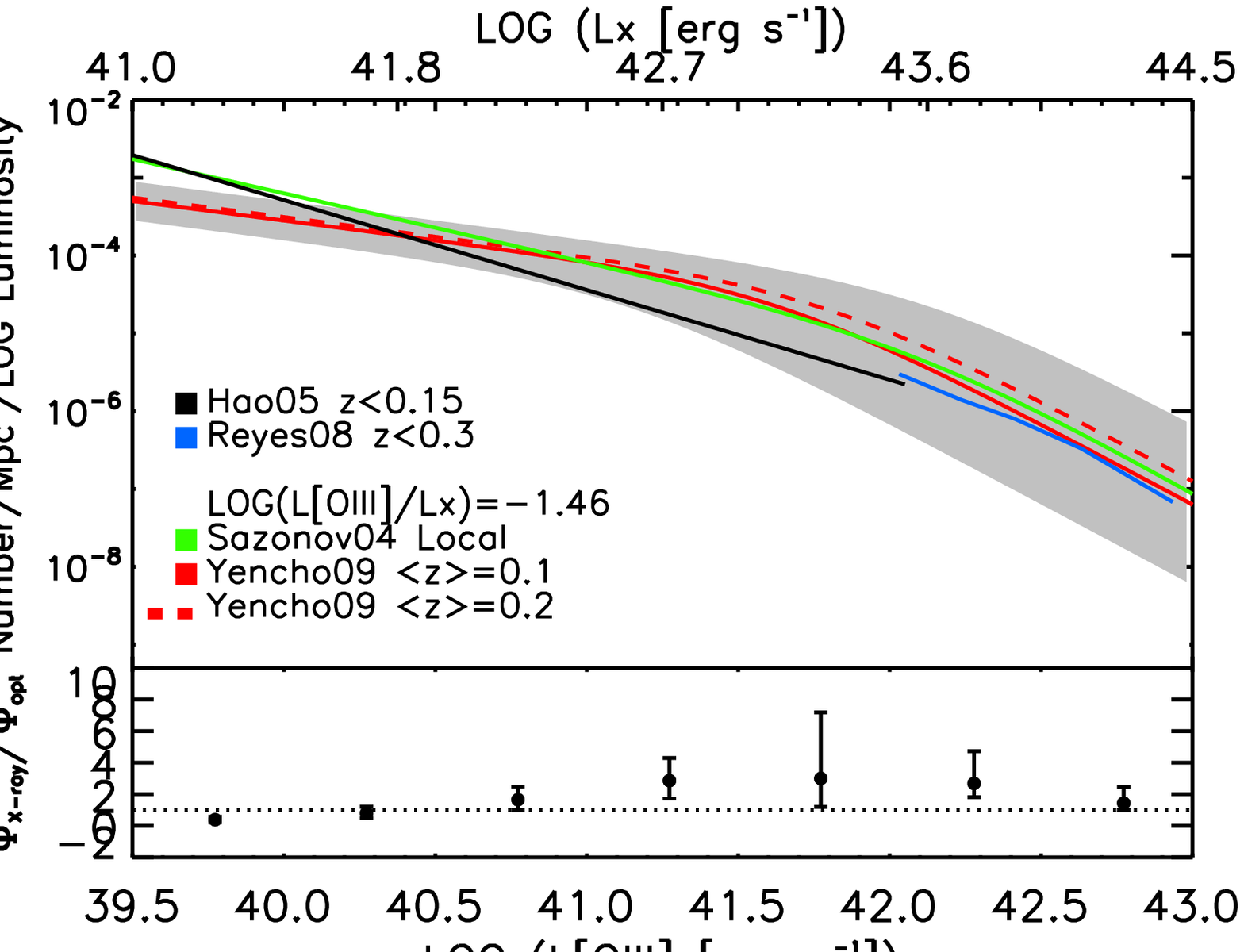} 
\vskip 0.1cm
\plotone{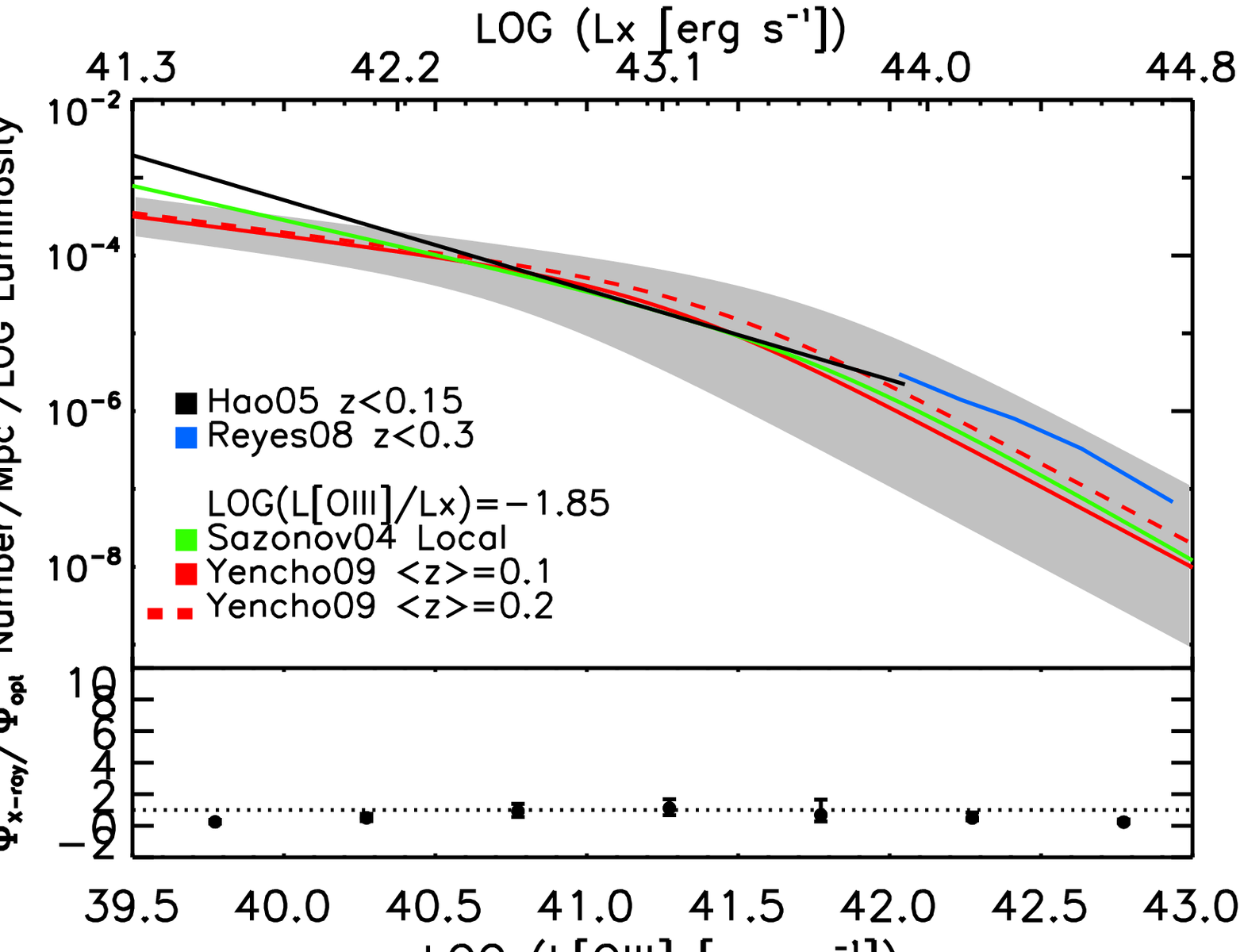} 
\caption{(a) Comparison of the X-ray and [OIII] LFs
(black curve---\citet{hao05} [OIII]
LF derived from the $z<0.15$ SDSS Seyfert~2s;
blue curve---\citet{reyes08} [OIII] LF
derived from the $z<0.3$ SDSS type~2 quasars; 
green curve---\citet{sazonov04} $3-20~\rm keV$ 
local {\em RXTE} LF; 
red curve---\citet{yencho09} $z=0.1$ 
extrapolation to their ILDE model best fit to the 
$0<z<1.2$ rest-frame $2-8~\rm keV$ distant+local LF;
red dashed curve---same as the red curve but
for a $z=0.2$ extrapolation). We aligned the X-ray LFs to the [OIII]
LFs using the BPT AGN-dominated non-BLAGNs
$\langle$\loglratio$\rangle=-1.06$. The gray shading shows the
$3~\sigma$ uncertainties for the transformed \citet{yencho09}
$z=0.1$ LF. Bottom panel shows the transformed \citet{yencho09}
X-ray LFs divided by the SDSS LFs. (b) Same as (a) but using 
the $z \le 0.5$ non-BLAGNs $\langle$\loglratio$\rangle=-1.46$, which 
excludes the upper limit values for the absorbers.
(c) Same as (a) but using the $z \le 0.5$ BLAGNs
$\langle$\loglratio$\rangle=-1.85$.
\label{lf}
}
\end{figure*}

In Figure~\ref{lf}(a) we show the SDSS [OIII] LF
for $z<0.15$ Seyfert~2s \citep[black curve]{hao05} 
and $z<0.3$ type~2 quasars \citep[blue curve]{reyes08}. 
Following \citet{reyes08}, we shift the \citet{hao05} luminosities 
by 0.14~dex to account for the difference between the spectrophotometric 
calibration scales. \citet{reyes08} stress that their type~2 quasar 
LF is a lower limit due to their selection criteria, the details 
of the spectroscopic target selection, and other factors.  
On this we show the \citet{sazonov04} $3-20~\rm keV$ 
LF based on their fit to the local {\em RXTE\/} sample (green curve). 
We also show the $z=0.1$
(red solid curve; this is the volume-weighted redshift 
that corresponds to the redshift range of the SDSS Seyfert~2s)
and $z=0.2$ (red dashed curve; this is the volume-weighted redshift
that corresponds to the redshift range of the SDSS type~2 quasars)
extrapolations of the \citet{yencho09} best fit
to their distant \emph{Chandra} plus local \emph{SWIFT BAT}
$2-8$~keV sample over the redshift interval $0<z<1.2$ for an
independent luminosity and density evolution (ILDE) model.
We have transformed the X-ray LFs to the [OIII] LFs using
$\langle$\loglratio$\rangle=-1.06$ 
(from the $z\le 0.5$ BPT AGN-dominated non-BLAGNs).  
The gray shading shows the $3~\sigma$
uncertainties on the transformation. Since neither \citet{hao05} nor
\citet{reyes08} apply a reddening correction to their [OIII]
luminosities, we use our extinction-uncorrected transformation. 

The bottom panel shows the transformed \citet{yencho09} LFs divided 
by the SDSS LFs. For $L_{\rm [OIII]}<10^{42}$~\ergss, we divide the
\citet{yencho09} $z=0.1$ LF by the \citet{hao05} $z<0.15$ LF. 
For $L_{\rm [OIII]}\ge 10^{42}$~\ergss, we divide the 
\citet{yencho09} $z=0.2$ LF by the \citet{reyes08} $z<0.3$ LF. 

In Figures~\ref{lf}(b) and (c) we instead did the transformations
using $\langle$\loglratio$\rangle=-1.46$ (from the $z \le 0.5$ non-BLAGNs,
excluding the upper limit values for the absorbers) 
and $\langle$\loglratio$\rangle=-1.85$ (from the $z \le 0.5$ BLAGNs).  

Despite the fact that there are strong selection effects in both 
the [OIII] and hard X-ray LFs (particularly in the way the
[OIII] LFs were constructed, since they contain only Seyfert~2s, and
due to the X-ray selection missing heavily X-ray absorbed,
Compton-thick sources),
we see reasonable agreement.  Thus, unlike \citet{heckman05}, we do 
not see any evidence for a substantial increase in the [OIII] LF
relative to the X-ray LF \citep[see also][]{georgantopoulos10}.

\section{Summary}
\label{secsummary}

We have used our highly optically spectroscopically complete
OPTX sample of $2-8$~keV X-ray selected AGNs to see how well an
empirical optical emission-line diagnostic diagram does in 
classifying the X-ray sources as AGNs.  We find that roughly
20\% of the X-ray AGNs that can be put on these diagrams are not
classified as AGNs by the
optical diagnostic diagram, if we use the \citet{kauffmann03}
empirical division between AGN and star formers. If instead we use the
\citet{kewley01} theoretical division, over half of our X-ray 
AGNs are not classified as AGNs.  
We show that both the BLAGNs and the non-BLAGNs in the OPTX sample 
exhibit a large (two orders in magnitude) dispersion in the ratio of
the [OIII] to hard X-ray 
luminosities, even after the application of reddening corrections 
to the [OIII] luminosities.  This confirms previous results from 
the literature.  The large observed spread in the luminosity
ratio leads us to postulate that the X-ray AGNs
that are misidentified by the optical diagnostic diagram as
star formers have low [OIII] luminosities due to the complexity 
of the structure of the narrow-line region.  If this is indeed
the case, then many of the ionizing photons from the AGN are 
not absorbed, and the [OIII] luminosity can only be used to predict
the X-ray luminosity to within a factor of $\sim 3$ (one sigma). While
\citet{melendez08}, \citet{rigby09}, and \citet{diamond-stanic09}
advocate [OIV] 26~$\mu$m as a more reliable indicator of ionizing
flux, it should be similarly affected by the complexity of the structure of
the narrow-line region.

\acknowledgements

We thank the anonymous referee for an extremely helpful report with
many useful suggestions for improving the paper. L.~T.~was supported
by a National Science Foundation Graduate Research
Fellowship and a Wisconsin Space Grant Consortium Graduate Fellowship
Award during portions of this work. We also gratefully acknowledge 
support from NSF grants AST~0708793, the University of
Wisconsin Research Committee with funds granted by the Wisconsin
Alumni Research Foundation, and the David and Lucile
Packard Foundation (A.~J.~B.). 

\bibliographystyle{apj}  
\bibliography{ms}

\begin{thebibliography}{63}
\expandafter\ifx\csname natexlab\endcsname\relax\def\natexlab#1{#1}\fi

\bibitem[{{Akylas} {et~al.}(2006){Akylas}, {Georgantopoulos}, {Georgakakis},
  {Kitsionas}, \& {Hatziminaoglou}}]{akylas06}
{Akylas}, A., {Georgantopoulos}, I., {Georgakakis}, A., {Kitsionas}, S., \&
  {Hatziminaoglou}, E. 2006, \aap, 459, 693

\bibitem[{{Alonso-Herrero} {et~al.}(1997){Alonso-Herrero}, {Ward}, \&
  {Kotilainen}}]{alonso-herrero97}
{Alonso-Herrero}, A., {Ward}, M.~J., \& {Kotilainen}, J.~K. 1997, \mnras, 288,
  977

\bibitem[{{Antonucci}(1993)}]{antonucci93}
{Antonucci}, R. 1993, \araa, 31, 473

\bibitem[{{Baldwin} {et~al.}(1981){Baldwin}, {Phillips}, \&
  {Terlevich}}]{baldwin81}
{Baldwin}, J.~A., {Phillips}, M.~M., \& {Terlevich}, R. 1981, \pasp, 93, 5
  (BPT)

\bibitem[{{Barger} {et~al.}(2002){Barger}, {Cowie}, {Brandt}, {Capak},
  {Garmire}, {Hornschemeier}, {Steffen}, \& {Wehner}}]{barger02}
{Barger}, A.~J., {Cowie}, L.~L., {Brandt}, W.~N., {Capak}, P., {Garmire},
  G.~P., {Hornschemeier}, A.~E., {Steffen}, A.~T., \& {Wehner}, E.~H. 2002,
  \aj, 124, 1839

\bibitem[{{Barger} {et~al.}(2005){Barger}, {Cowie}, {Mushotzky}, {Yang},
  {Wang}, {Steffen}, \& {Capak}}]{barger05}
{Barger}, A.~J., {Cowie}, L.~L., {Mushotzky}, R.~F., {Yang}, Y., {Wang}, W.-H.,
  {Steffen}, A.~T., \& {Capak}, P. 2005, \aj, 129, 578

\bibitem[{{Bassani} {et~al.}(1999){Bassani}, {Dadina}, {Maiolino}, {Salvati},
  {Risaliti}, {della Ceca}, {Matt}, \& {Zamorani}}]{bassani99}
{Bassani}, L., {Dadina}, M., {Maiolino}, R., {Salvati}, M., {Risaliti}, G.,
  {della Ceca}, R., {Matt}, G., \& {Zamorani}, G. 1999, \apjs, 121, 473

\bibitem[{{Beckmann} {et~al.}(2006){Beckmann}, {Soldi}, {Shrader}, {Gehrels},
  \& {Produit}}]{beckmann06}
{Beckmann}, V., {Soldi}, S., {Shrader}, C.~R., {Gehrels}, N., \& {Produit}, N.
  2006, \apj, 652, 126

\bibitem[{{Blanton} \& {Roweis}(2007)}]{blanton07}
{Blanton}, M.~R. \& {Roweis}, S. 2007, \aj, 133, 734

\bibitem[{{Bongiorno} {et~al.}(2010){Bongiorno}, {Mignoli}, {Zamorani},
  {Lamareille}, {Lanzuisi}, {Miyaji}, {Bolzonella}, {Carollo}, {Contini},
  {Kneib}, {Le F{\`e}vre}, {Lilly}, {Mainieri}, {Renzini}, {Scodeggio},
  {Bardelli}, {Brusa}, {Caputi}, {Civano}, {Coppa}, {Cucciati}, {de La Torre},
  {de Ravel}, {Franzetti}, {Garilli}, {Halliday}, {Hasinger}, {Koekemoer},
  {Iovino}, {Kampczyk}, {Knobel}, {Kova{\v c}}, {Le Borgne}, {Le Brun},
  {Maier}, {Merloni}, {Nair}, {Pello}, {Peng}, {Perez Montero}, {Ricciardelli},
  {Salvato}, {Silverman}, {Tanaka}, {Tasca}, {Tresse}, {Vergani}, {Zucca},
  {Abbas}, {Bottini}, {Cappi}, {Cassata}, {Cimatti}, {Guzzo}, {Leauthaud},
  {Maccagni}, {Marinoni}, {McCracken}, {Memeo}, {Meneux}, {Oesch}, {Porciani},
  {Pozzetti}, \& {Scaramella}}]{bongiorno10}
{Bongiorno}, A., {Mignoli}, M., {Zamorani}, G., {Lamareille}, F., {Lanzuisi},
  G., {Miyaji}, T., {Bolzonella}, M., {Carollo}, C.~M., {Contini}, T., {Kneib},
  J.~P., {Le F{\`e}vre}, O., {Lilly}, S.~J., {Mainieri}, V., {Renzini}, A.,
  {Scodeggio}, M., {Bardelli}, S., {Brusa}, M., {Caputi}, K., {Civano}, F.,
  {Coppa}, G., {Cucciati}, O., {de La Torre}, S., {de Ravel}, L., {Franzetti},
  P., {Garilli}, B., {Halliday}, C., {Hasinger}, G., {Koekemoer}, A.~M.,
  {Iovino}, A., {Kampczyk}, P., {Knobel}, C., {Kova{\v c}}, K., {Le Borgne},
  J., {Le Brun}, V., {Maier}, C., {Merloni}, A., {Nair}, P., {Pello}, R.,
  {Peng}, Y., {Perez Montero}, E., {Ricciardelli}, E., {Salvato}, M.,
  {Silverman}, J., {Tanaka}, M., {Tasca}, L., {Tresse}, L., {Vergani}, D.,
  {Zucca}, E., {Abbas}, U., {Bottini}, D., {Cappi}, A., {Cassata}, P.,
  {Cimatti}, A., {Guzzo}, L., {Leauthaud}, A., {Maccagni}, D., {Marinoni}, C.,
  {McCracken}, H.~J., {Memeo}, P., {Meneux}, B., {Oesch}, P., {Porciani}, C.,
  {Pozzetti}, L., \& {Scaramella}, R. 2010, \aap, 510, A56

\bibitem[{{Bruzual} \& {Charlot}(2003)}]{bruzual03}
{Bruzual}, G. \& {Charlot}, S. 2003, \mnras, 344, 1000

\bibitem[{{Cocchia} {et~al.}(2007){Cocchia}, {Fiore}, {Vignali}, {Mignoli},
  {Brusa}, {Comastri}, {Feruglio}, {Baldi}, {Carangelo}, {Ciliegi}, {D'Elia},
  {La Franca}, {Maiolino}, {Matt}, {Molendi}, {Perola}, \&
  {Puccetti}}]{cocchia07}
{Cocchia}, F., {Fiore}, F., {Vignali}, C., {Mignoli}, M., {Brusa}, M.,
  {Comastri}, A., {Feruglio}, C., {Baldi}, A., {Carangelo}, N., {Ciliegi}, P.,
  {D'Elia}, V., {La Franca}, F., {Maiolino}, R., {Matt}, G., {Molendi}, S.,
  {Perola}, G.~C., \& {Puccetti}, S. 2007, \aap, 466, 31

\bibitem[{{Comastri}(2004)}]{comastri04}
{Comastri}, A. 2004, in Astrophysics and Space Science Library, Vol. 308,
  Supermassive Black Holes in the Distant Universe, ed. {A.~J.~Barger}, 245

\bibitem[{{Cowie} \& {Barger}(2008)}]{cowie08}
{Cowie}, L.~L. \& {Barger}, A.~J. 2008, \apj, 686, 72

\bibitem[{{Dahari} \& {De Robertis}(1988)}]{dahari88}
{Dahari}, O. \& {De Robertis}, M.~M. 1988, \apjs, 67, 249

\bibitem[{{Diamond-Stanic} {et~al.}(2009){Diamond-Stanic}, {Rieke}, \&
  {Rigby}}]{diamond-stanic09}
{Diamond-Stanic}, A.~M., {Rieke}, G.~H., \& {Rigby}, J.~R. 2009, \apj, 698, 623

\bibitem[{{Faber} {et~al.}(2003){Faber}, {Phillips}, {Kibrick}, {Alcott},
  {Allen}, {Burrous}, {Cantrall}, {Clarke}, {Coil}, {Cowley}, {Davis}, {Deich},
  {Dietsch}, {Gilmore}, {Harper}, {Hilyard}, {Lewis}, {McVeigh}, {Newman},
  {Osborne}, {Schiavon}, {Stover}, {Tucker}, {Wallace}, {Wei}, {Wirth}, \&
  {Wright}}]{faber03}
{Faber}, S.~M., {Phillips}, A.~C., {Kibrick}, R.~I., {Alcott}, B., {Allen},
  S.~L., {Burrous}, J., {Cantrall}, T., {Clarke}, D., {Coil}, A.~L., {Cowley},
  D.~J., {Davis}, M., {Deich}, W.~T.~S., {Dietsch}, K., {Gilmore}, D.~K.,
  {Harper}, C.~A., {Hilyard}, D.~F., {Lewis}, J.~P., {McVeigh}, M., {Newman},
  J., {Osborne}, J., {Schiavon}, R., {Stover}, R.~J., {Tucker}, D., {Wallace},
  V., {Wei}, M., {Wirth}, G., \& {Wright}, C.~A. 2003, Proc. SPIE, 4841, 1657

\bibitem[{{Ferland} \& {Netzer}(1983)}]{ferland83}
{Ferland}, G.~J. \& {Netzer}, H. 1983, \apj, 264, 105

\bibitem[{{Ferland} \& {Osterbrock}(1986)}]{ferland86}
{Ferland}, G.~J. \& {Osterbrock}, D.~E. 1986, \apj, 300, 658

\bibitem[{{Ferruit} {et~al.}(1999){Ferruit}, {Wilson}, {Whittle}, {Simpson},
  {Mulchaey}, \& {Ferland}}]{ferruit99}
{Ferruit}, P., {Wilson}, A.~S., {Whittle}, M., {Simpson}, C., {Mulchaey},
  J.~S., \& {Ferland}, G.~J. 1999, \apj, 523, 147

\bibitem[{{Georgantopoulos} \& {Akylas}(2010)}]{georgantopoulos10}
{Georgantopoulos}, I. \& {Akylas}, A. 2010, \aap, 509, A26

\bibitem[{{Guainazzi} {et~al.}(2005){Guainazzi}, {Matt}, \&
  {Perola}}]{guainazzi05}
{Guainazzi}, M., {Matt}, G., \& {Perola}, G.~C. 2005, \aap, 444, 119

\bibitem[{{Hao} {et~al.}(2005){Hao}, {Strauss}, {Fan}, {Tremonti}, {Schlegel},
  {Heckman}, {Kauffmann}, {Blanton}, {Gunn}, {Hall}, {Ivezi{\'c}}, {Knapp},
  {Krolik}, {Lupton}, {Richards}, {Schneider}, {Strateva}, {Zakamska},
  {Brinkmann}, \& {Szokoly}}]{hao05}
{Hao}, L., {Strauss}, M.~A., {Fan}, X., {Tremonti}, C.~A., {Schlegel}, D.~J.,
  {Heckman}, T.~M., {Kauffmann}, G., {Blanton}, M.~R., {Gunn}, J.~E., {Hall},
  P.~B., {Ivezi{\'c}}, {\v Z}., {Knapp}, G.~R., {Krolik}, J.~H., {Lupton},
  R.~H., {Richards}, G.~T., {Schneider}, D.~P., {Strateva}, I.~V., {Zakamska},
  N.~L., {Brinkmann}, J., \& {Szokoly}, G.~P. 2005, \aj, 129, 1795

\bibitem[{{Hasinger}(2008)}]{hasinger08}
{Hasinger}, G. 2008, \aap, 490, 905

\bibitem[{{Heckman} {et~al.}(2004){Heckman}, {Kauffmann}, {Brinchmann},
  {Charlot}, {Tremonti}, \& {White}}]{heckman04}
{Heckman}, T.~M., {Kauffmann}, G., {Brinchmann}, J., {Charlot}, S., {Tremonti},
  C., \& {White}, S.~D.~M. 2004, \apj, 613, 109

\bibitem[{{Heckman} {et~al.}(2005){Heckman}, {Ptak}, {Hornschemeier}, \&
  {Kauffmann}}]{heckman05}
{Heckman}, T.~M., {Ptak}, A., {Hornschemeier}, A., \& {Kauffmann}, G. 2005,
  \apj, 634, 161

\bibitem[{{Kakazu} {et~al.}(2007){Kakazu}, {Cowie}, \& {Hu}}]{kakazu07}
{Kakazu}, Y., {Cowie}, L.~L., \& {Hu}, E.~M. 2007, \apj, 668, 853

\bibitem[{{Kauffmann} {et~al.}(2003){Kauffmann}, {Heckman}, {Tremonti},
  {Brinchmann}, {Charlot}, {White}, {Ridgway}, {Brinkmann}, {Fukugita}, {Hall},
  {Ivezi{\'c}}, {Richards}, \& {Schneider}}]{kauffmann03}
{Kauffmann}, G., {Heckman}, T.~M., {Tremonti}, C., {Brinchmann}, J., {Charlot},
  S., {White}, S.~D.~M., {Ridgway}, S.~E., {Brinkmann}, J., {Fukugita}, M.,
  {Hall}, P.~B., {Ivezi{\'c}}, {\v Z}., {Richards}, G.~T., \& {Schneider},
  D.~P. 2003, \mnras, 346, 1055

\bibitem[{{Kewley} {et~al.}(2001){Kewley}, {Dopita}, {Sutherland}, {Heisler},
  \& {Trevena}}]{kewley01}
{Kewley}, L.~J., {Dopita}, M.~A., {Sutherland}, R.~S., {Heisler}, C.~A., \&
  {Trevena}, J. 2001, \apj, 556, 121

\bibitem[{{Kewley} {et~al.}(2006){Kewley}, {Groves}, {Kauffmann}, \&
  {Heckman}}]{kewley06}
{Kewley}, L.~J., {Groves}, B., {Kauffmann}, G., \& {Heckman}, T. 2006, \mnras,
  372, 961

\bibitem[{{La Franca} {et~al.}(2005){La Franca}, {Fiore}, {Comastri}, {Perola},
  {Sacchi}, {Brusa}, {Cocchia}, {Feruglio}, {Matt}, {Vignali}, {Carangelo},
  {Ciliegi}, {Lamastra}, {Maiolino}, {Mignoli}, {Molendi}, \&
  {Puccetti}}]{lafranca05}
{La Franca}, F., {Fiore}, F., {Comastri}, A., {Perola}, G.~C., {Sacchi}, N.,
  {Brusa}, M., {Cocchia}, F., {Feruglio}, C., {Matt}, G., {Vignali}, C.,
  {Carangelo}, N., {Ciliegi}, P., {Lamastra}, A., {Maiolino}, R., {Mignoli},
  M., {Molendi}, S., \& {Puccetti}, S. 2005, \apj, 635, 864

\bibitem[{{La Massa} {et~al.}(2009){La Massa}, {Heckman}, {Ptak},
  {Hornschemeier}, {Martins}, {Sonnentrucker}, \& {Tremonti}}]{lamassa09}
{La Massa}, S.~M., {Heckman}, T.~M., {Ptak}, A., {Hornschemeier}, A.,
  {Martins}, L., {Sonnentrucker}, P., \& {Tremonti}, C. 2009, \apj, 705, 568

\bibitem[{{Laurent-Muehleisen} {et~al.}(1998){Laurent-Muehleisen}, {Kollgaard},
  {Ciardullo}, {Feigelson}, {Brinkmann}, \& {Siebert}}]{laurent-muehleisen98}
{Laurent-Muehleisen}, S.~A., {Kollgaard}, R.~I., {Ciardullo}, R., {Feigelson},
  E.~D., {Brinkmann}, W., \& {Siebert}, J. 1998, \apjs, 118, 127

\bibitem[{{Lawrence} \& {Elvis}(1982)}]{lawrence82}
{Lawrence}, A. \& {Elvis}, M. 1982, \apj, 256, 410

\bibitem[{{Magnelli} {et~al.}(2009){Magnelli}, {Elbaz}, {Chary}, {Dickinson},
  {Le Borgne}, {Frayer}, \& {Willmer}}]{magnelli09}
{Magnelli}, B., {Elbaz}, D., {Chary}, R.~R., {Dickinson}, M., {Le Borgne}, D.,
  {Frayer}, D.~T., \& {Willmer}, C.~N.~A. 2009, \aap, 496, 57

\bibitem[{{Mel{\'e}ndez} {et~al.}(2008){Mel{\'e}ndez}, {Kraemer}, {Armentrout},
  {Deo}, {Crenshaw}, {Schmitt}, {Mushotzky}, {Tueller}, {Markwardt}, \&
  {Winter}}]{melendez08}
{Mel{\'e}ndez}, M., {Kraemer}, S.~B., {Armentrout}, B.~K., {Deo}, R.~P.,
  {Crenshaw}, D.~M., {Schmitt}, H.~R., {Mushotzky}, R.~F., {Tueller}, J.,
  {Markwardt}, C.~B., \& {Winter}, L. 2008, \apj, 682, 94

\bibitem[{{Moran} {et~al.}(1999){Moran}, {Lehnert}, \& {Helfand}}]{moran99}
{Moran}, E.~C., {Lehnert}, M.~D., \& {Helfand}, D.~J. 1999, \apj, 526, 649

\bibitem[{{Mulchaey} {et~al.}(1994){Mulchaey}, {Koratkar}, {Ward}, {Wilson},
  {Whittle}, {Antonucci}, {Kinney}, \& {Hurt}}]{mulchaey94}
{Mulchaey}, J.~S., {Koratkar}, A., {Ward}, M.~J., {Wilson}, A.~S., {Whittle},
  M., {Antonucci}, R.~R.~J., {Kinney}, A.~L., \& {Hurt}, T. 1994, \apj, 436,
  586

\bibitem[{{Netzer} {et~al.}(2006){Netzer}, {Mainieri}, {Rosati}, \&
  {Trakhtenbrot}}]{netzer06}
{Netzer}, H., {Mainieri}, V., {Rosati}, P., \& {Trakhtenbrot}, B. 2006, \aap,
  453, 525

\bibitem[{{Osterbrock} \& {Martel}(1993)}]{osterbrock93}
{Osterbrock}, D.~E. \& {Martel}, A. 1993, \apj, 414, 552

\bibitem[{{Osterbrock} \& {Pogge}(1985)}]{osterbrock85}
{Osterbrock}, D.~E. \& {Pogge}, R.~W. 1985, \apj, 297, 166

\bibitem[{{Ptak} {et~al.}(2006){Ptak}, {Zakamska}, {Strauss}, {Krolik},
  {Heckman}, {Schneider}, \& {Brinkmann}}]{ptak06}
{Ptak}, A., {Zakamska}, N.~L., {Strauss}, M.~A., {Krolik}, J.~H., {Heckman},
  T.~M., {Schneider}, D.~P., \& {Brinkmann}, J. 2006, \apj, 637, 147

\bibitem[{{Ranalli} {et~al.}(2003){Ranalli}, {Comastri}, \&
  {Setti}}]{ranalli03}
{Ranalli}, P., {Comastri}, A., \& {Setti}, G. 2003, \aap, 399, 39

\bibitem[{{Reyes} {et~al.}(2008){Reyes}, {Zakamska}, {Strauss}, {Green},
  {Krolik}, {Shen}, {Richards}, {Anderson}, \& {Schneider}}]{reyes08}
{Reyes}, R., {Zakamska}, N.~L., {Strauss}, M.~A., {Green}, J., {Krolik}, J.~H.,
  {Shen}, Y., {Richards}, G.~T., {Anderson}, S.~F., \& {Schneider}, D.~P. 2008,
  \aj, 136, 2373

\bibitem[{{Rigby} {et~al.}(2009){Rigby}, {Diamond-Stanic}, \&
  {Aniano}}]{rigby09}
{Rigby}, J.~R., {Diamond-Stanic}, A.~M., \& {Aniano}, G. 2009, \apj, 700, 1878

\bibitem[{{Sazonov} {et~al.}(2007){Sazonov}, {Revnivtsev}, {Krivonos},
  {Churazov}, \& {Sunyaev}}]{sazonov07}
{Sazonov}, S., {Revnivtsev}, M., {Krivonos}, R., {Churazov}, E., \& {Sunyaev},
  R. 2007, \aap, 462, 57

\bibitem[{{Sazonov} \& {Revnivtsev}(2004)}]{sazonov04}
{Sazonov}, S.~Y. \& {Revnivtsev}, M.~G. 2004, \aap, 423, 469

\bibitem[{{Silverman} {et~al.}(2008){Silverman}, {Green}, {Barkhouse}, {Kim},
  {Kim}, {Wilkes}, {Cameron}, {Hasinger}, {Jannuzi}, {Smith}, {Smith}, \&
  {Tananbaum}}]{silverman08}
{Silverman}, J.~D., {Green}, P.~J., {Barkhouse}, W.~A., {Kim}, D.-W., {Kim},
  M., {Wilkes}, B.~J., {Cameron}, R.~A., {Hasinger}, G., {Jannuzi}, B.~T.,
  {Smith}, M.~G., {Smith}, P.~S., \& {Tananbaum}, H. 2008, \apj, 679, 118

\bibitem[{{Simpson}(2005)}]{simpson05}
{Simpson}, C. 2005, \mnras, 360, 565

\bibitem[{{Stasi{\'n}ska} {et~al.}(2006){Stasi{\'n}ska}, {Cid Fernandes},
  {Mateus}, {Sodr{\'e}}, \& {Asari}}]{stasinska06}
{Stasi{\'n}ska}, G., {Cid Fernandes}, R., {Mateus}, A., {Sodr{\'e}}, L., \&
  {Asari}, N.~V. 2006, \mnras, 371, 972

\bibitem[{{Steffen} {et~al.}(2003){Steffen}, {Barger}, {Cowie}, {Mushotzky}, \&
  {Yang}}]{steffen03}
{Steffen}, A.~T., {Barger}, A.~J., {Cowie}, L.~L., {Mushotzky}, R.~F., \&
  {Yang}, Y. 2003, \apjl, 596, L23

\bibitem[{{Szokoly} {et~al.}(2004){Szokoly}, {Bergeron}, {Hasinger}, {Lehmann},
  {Kewley}, {Mainieri}, {Nonino}, {Rosati}, {Giacconi}, {Gilli}, {Gilmozzi},
  {Norman}, {Romaniello}, {Schreier}, {Tozzi}, {Wang}, {Zheng}, \&
  {Zirm}}]{szokoly04}
{Szokoly}, G.~P., {Bergeron}, J., {Hasinger}, G., {Lehmann}, I., {Kewley}, L.,
  {Mainieri}, V., {Nonino}, M., {Rosati}, P., {Giacconi}, R., {Gilli}, R.,
  {Gilmozzi}, R., {Norman}, C., {Romaniello}, M., {Schreier}, E., {Tozzi}, P.,
  {Wang}, J.~X., {Zheng}, W., \& {Zirm}, A. 2004, \apjs, 155, 271

\bibitem[{{Tremonti} {et~al.}(2004){Tremonti}, {Heckman}, {Kauffmann},
  {Brinchmann}, {Charlot}, {White}, {Seibert}, {Peng}, {Schlegel}, {Uomoto},
  {Fukugita}, \& {Brinkmann}}]{tremonti04}
{Tremonti}, C.~A., {Heckman}, T.~M., {Kauffmann}, G., {Brinchmann}, J.,
  {Charlot}, S., {White}, S.~D.~M., {Seibert}, M., {Peng}, E.~W., {Schlegel},
  D.~J., {Uomoto}, A., {Fukugita}, M., \& {Brinkmann}, J. 2004, \apj, 613, 898

\bibitem[{{Trouille} {et~al.}(2008){Trouille}, {Barger}, {Cowie}, {Yang}, \&
  {Mushotzky}}]{trouille08}
{Trouille}, L., {Barger}, A.~J., {Cowie}, L.~L., {Yang}, Y., \& {Mushotzky},
  R.~F. 2008, \apjs, 179, 1

\bibitem[{{Trouille} {et~al.}(2009){Trouille}, {Barger}, {Cowie}, {Yang}, \&
  {Mushotzky}}]{trouille09}
---. 2009, \apj, 703, 2160

\bibitem[{{Ueda} {et~al.}(2003){Ueda}, {Akiyama}, {Ohta}, \& {Miyaji}}]{ueda03}
{Ueda}, Y., {Akiyama}, M., {Ohta}, K., \& {Miyaji}, T. 2003, \apj, 598, 886

\bibitem[{{Veilleux} \& {Osterbrock}(1987)}]{veilleux87}
{Veilleux}, S. \& {Osterbrock}, D.~E. 1987, \apjs, 63, 295

\bibitem[{{Weedman}(1977)}]{weedman77}
{Weedman}, D.~W. 1977, \araa, 15, 69

\bibitem[{{Weedman}(1978)}]{weedman78}
---. 1978, \mnras, 184, 11P

\bibitem[{{Winter} {et~al.}(2009){Winter}, {Mushotzky}, {Reynolds}, \&
  {Tueller}}]{winter09}
{Winter}, L.~M., {Mushotzky}, R.~F., {Reynolds}, C.~S., \& {Tueller}, J. 2009,
  \apj, 690, 1322

\bibitem[{{Yencho} {et~al.}(2009){Yencho}, {Barger}, {Trouille}, \&
  {Winter}}]{yencho09}
{Yencho}, B., {Barger}, A.~J., {Trouille}, L., \& {Winter}, L.~M. 2009, \apj,
  698, 380

\bibitem[{{Zakamska} {et~al.}(2003){Zakamska}, {Strauss}, {Krolik}, {Collinge},
  {Hall}, {Hao}, {Heckman}, {Ivezi{\'c}}, {Richards}, {Schlegel}, {Schneider},
  {Strateva}, {Vanden Berk}, {Anderson}, \& {Brinkmann}}]{zakamska03}
{Zakamska}, N.~L., {Strauss}, M.~A., {Krolik}, J.~H., {Collinge}, M.~J.,
  {Hall}, P.~B., {Hao}, L., {Heckman}, T.~M., {Ivezi{\'c}}, {\v Z}.,
  {Richards}, G.~T., {Schlegel}, D.~J., {Schneider}, D.~P., {Strateva}, I.,
  {Vanden Berk}, D.~E., {Anderson}, S.~F., \& {Brinkmann}, J. 2003, \aj, 126,
  2125

\bibitem[{{Zezas} {et~al.}(1998){Zezas}, {Georgantopoulos}, \&
  {Ward}}]{zezas98}
{Zezas}, A.~L., {Georgantopoulos}, I., \& {Ward}, M.~J. 1998, \mnras, 301, 915

\end{thebibliography}

\end{document}